\documentclass[aps,prd,showpacs,notitlepage,eqsecnum,
superscriptaddress]{revtex4-1}

\usepackage[centertags]{amsmath}
\usepackage{amssymb}
\usepackage{latexsym}
\usepackage{enumerate}
\usepackage{graphicx}
\usepackage{mathrsfs}
\usepackage{hyperref}
\usepackage{stmaryrd}
\usepackage{import}
\usepackage{tensor}
\usepackage{color}
\usepackage{bm}
\usepackage{multirow}
\usepackage{breakurl}
\usepackage{float}
\usepackage{placeins}

\newcommand{\bi}{\begin{itemize}}
\newcommand{\ei}{\end{itemize}}
\newcommand{\be}{\begin{equation}}
\newcommand{\ee}{\end{equation}}

\renewcommand{\l}{\left(}
\renewcommand{\r}{\right)}
\renewcommand{\a}{\alpha}

\renewcommand{\d}{\delta}
\newcommand{\D}{\Delta}

\newcommand{\La}{\Lambda}
\newcommand{\la}{\lambda}
\renewcommand{\O}{\Omega}
\renewcommand{\o}{\omega}
\renewcommand{\th}{\theta}

\newcommand{\q}{\quad}

\newcommand{\vp}{\varphi}

\newcommand{\pa}{\partial}

\begin{document}

\title{Analytic 
self-force calculations
in the post-Newtonian regime:\\ 
eccentric orbits on a Schwarzschild background
}

\author{Seth Hopper}
\affiliation{School of Mathematics and Statistics and Complex \& Adaptive Systems 
Laboratory, University College Dublin, Belfield, Dublin 4, Ireland}
\affiliation{CENTRA, Dept. de F\'{i}sica, Instituto Superior T\'{e}cnico – IST,
Av. Rovisco Pais 1, 1049, Lisboa, Portugal}
\author{Chris Kavanagh}
\affiliation{School of Mathematics and Statistics and Complex \& Adaptive Systems 
Laboratory, University College Dublin, Belfield, Dublin 4, Ireland}
\author{Adrian C. Ottewill}
\affiliation{School of Mathematics and Statistics and Complex \& Adaptive Systems 
Laboratory, University College Dublin, Belfield, Dublin 4, Ireland}

\begin{abstract}
We present a method for solving the first-order
field equations in a post-Newtonian (PN) expansion.
Our calculations generalize work of Bini and Damour
and subsequently Kavanagh et al., to
consider eccentric orbits on a Schwarzschild background.
We derive expressions for the 
retarded metric perturbation at the location of the particle for 
all $\ell$-modes. We find that, despite first appearances, the Regge-Wheeler
gauge metric perturbation is $C^0$ at the particle for all $\ell$.
As a first use of our solutions, we compute the gauge-invariant quantity
$\langle U \rangle$ through 4PN while simultaneously expanding
in eccentricity through $e^{10}$.
By anticipating the $e\to 1$ singular 
behavior at each PN order, we greatly
improve the accuracy of our results for large $e$.
We use $\langle U \rangle$ to find 4PN contributions
to the effective one body potential $\hat Q$ through $e^{10}$
and at linear order in the mass-ratio.
\end{abstract}


\maketitle

\section{Introduction}
\label{sec:intro}

Recent years have seen a large amount of research in the regime 
of overlap between two complementary approaches
to the general relativistic two body problem: 
gravitational self-force (GSF)  and
post-Newtonian (PN) theory. 
GSF calculations are made within the context of
black hole perturbation theory, an expansion in the 
mass-ratio $q \equiv \mu /M$ of the two bodies, which is valid for all speeds.
PN theory, on the other hand, is an expansion in small velocities
(or equivalently, large separations),
but is valid for any $q$. In the area of parameter space where the
two theories overlap, they can check one another and also be used to
compute previously unknown parameters.

Much of the recent GSF/PN work has been performed 
with the eventual goal of detecting
gravitational waves. With Advanced LIGO \cite{LIGO} now performing science runs,
the need for accurate waveforms is immediate. One very effective framework for
producing such waveforms is the effective one body (EOB) model \cite{BuonDamo99}.
Careful GSF calculations, when performed in
the PN regime, can then be used to 
provide information on the PN behavior of EOB potentials.

This calibration is possible only because
gauge-invariant quantities can be computed separately using the GSF and PN theory. 
The first such invariant was the ``redshift invariant'' $u^t$,
originally suggested by Detweiler \cite{Detw05},
who then computed $u^t$ numerically with a GSF code and compared it to its PN
value through third PN order \cite{Detw08}. 
Subsequently, a number of papers used numerical techniques to
compute $u^t$ to ever higher precision, and fit out previously unknown PN parameters
at ever higher order 
\cite{BlanETC09,BlanETC10,ShahFrieWhit14,BlanFayeWhit14a,BlanFayeWhit14b}.
Work by Shah et al.~\cite{ShahFrieWhit14}, in particular, 
opened a new avenue to obtaining 
high-order PN parameters in the linear-in-$q$ limit. 
Rather than solve the first-order field equations through numerical integration, 
they employed  the function expansion method of Mano, Suzuki and Takasugi (MST)
\cite{ManoSuzuTaka96a,ManoSuzuTaka96b}.
Combining MST with computer algebra software like \emph{Mathematica}, one can solve
the field equations to hundreds or even thousands of digits
when considering large radii orbits.

In addition to $u^t$, other local, circular orbit gauge invariants have been calculated.
Barack and Sago \cite{BaraSago09} computed the shift to the innermost stable
circular orbit due to the GSF (which subsequently was used by Akcay et al.~to inform 
EOB in Ref.~\cite{AkcaETC12}). Then, inspired by work of Harte \cite{Hart12},
Dolan et al.~\cite{DolaETC14a} computed the so-called ``spin-invariant'' $\psi$,
which measures the GSF effect on geodetic spin precession. 
Subsequently (as suggested in Ref.~\cite{BiniDamo14b}), 
Dolan et al.~\cite{DolaETC14b} computed higher-order
``tidal invariants'', followed by Nolan et al.~\cite{NolaETC15} computing
``octupolar invariants''.

All of those calculations were restricted to circular orbits, which
are much simpler computationally than eccentric orbits.
There are a number of reasons why. First, because circular orbits posses only
one harmonic for each $\ell m$ mode, the field equations at that mode
reduce from a set of 1+1 partial differential equations to
ordinary differential equations. When considering eccentric orbits, one must
either solve the 1+1 equations directly in the time domain (TD) as an initial
value problem, or stay in the frequency domain (FD), wherein the
number of harmonics goes from only one to a countably infinite set 
(although in practice, of course, this is truncated based on some accuracy 
criterion). Additionally, finding the particular solution in the FD is no
longer a simple matching, but rather requires an integration over
one period of the source's libration. 

Despite these difficulties, gauge invariants have been computed
for eccentric orbits as well.
As suggested first by Damour \cite{Damo10}, Barack et al.~\cite{BaraDamoSago10}
computed the GSF effect on the precession rate of slightly-eccentric orbits
around a Schwarzschild black hole.
Then, Barack and Sago \cite{BaraSago11} generalized Detweiler's $u^t$ 
to generic bound orbits on a Schwarzschild background by
proper time averaging it over the perturbed orbit to form $\langle U \rangle$.
In that same work, they computed the periapsis advance of eccentric orbits,
another gauge invariant. The work in both Refs.~\cite{BaraDamoSago10} 
and \cite{BaraSago11} employed a TD code in the strong field regime. 
Since then, $\langle U \rangle$ has been thoroughly 
examined in the weaker field by Akcay et al.~\cite{AkcaETC15} using a FD code and 
direct analytic PN calculations through 3PN. In addition, 
van de Meent and Shah
have computed $\langle U \rangle$ for equatorial orbits 
on a Kerr background \cite{VandShah15} for the first time.
Along with providing checks between PN and GSF, these calculations have
served as important internal consistency checks for GSF, wherein 
$\langle U \rangle$ has been found in Lorenz, radiation,
and Regge-Wheeler gauges.

To add to the numerical approaches merging PN and GSF, there has been
ongoing analytic work. 
Indeed, the combined use of black hole perturbation theory
and PN theory has an extensive history largely inspired by the original 
MST papers \cite{ManoSuzuTaka96a,ManoSuzuTaka96b}. Sago, Nakano, Hikida,
Fujita (and many others),
\cite{SagoNakaSasa03,NakaSagoSasa03,HikiETC04,HikiETC05,GanzETC07,Fuji12,Fuji15,SagoFuji15} 
have used PN expansions to compute fluxes, waveforms and the GSF
for a variety of orbits on
both Schwarzschild and Kerr backgrounds. 
More recently, in a series of papers 
Bini and Damour \cite{BiniDamo13,BiniDamo14a,BiniDamo14b,BiniDamo14c} 
have used these methods to analytically find local gauge invariants.
They used MST to compute $u^t$, 
$\psi$, as well as tidal invariants in the linear-in-$q$ limit,
with a focus on EOB calibration.
Since then, Kavanagh et al.~\cite{KavaOtteWard15} have built upon
the Bini and Damour approach and computed $u^t$, $\psi$, 
and the tidal invariants, to 21.5PN, the current state-of-the-art.

In this work we present a method for computing 
GSF quantities sourced by eccentric orbits around a Schwarzschild black hole
through use of an analytic PN expansion. Our method extends the circular orbit
work of \cite{BiniDamo13,KavaOtteWard15} by performing an expansion in small eccentricity at each PN order.
Our calculations are performed in Regge-Wheeler \cite{ReggWhee57} gauge for all 
$\ell \ge 2$ and make use of Zerilli's \cite{Zeri70} 
analytic solutions for $\ell = 0, 1$ 
(although with a slight gauge transformation to the monopole). 
We collectively refer to this as Regge-Wheeler-Zerilli (RWZ) gauge.
We obtain expressions for the retarded metric perturbation (MP) and its first
derivatives for $\ell$ modes as a function of the particle's position.
Significantly, in our final expressions, we identify poles in the small-$e$
expansion. After factoring out singular-in-$e$ terms at each PN order, we greatly
improve the accuracy of our results for large eccentricities.
As a use of our solutions, we compute $\langle U \rangle$,
confirming the numeric 4PN predictions of 
Ref.~\cite{VandShah15} and provide new PN parameters for 4PN coefficients
through $e^{10}$. 
We note similar 
concurrent work \cite{Hopp15} of Bini et al.~\cite{BiniDamoGera15}, who worked
to 6.5PN and $e^2$.

The paper is arranged as follows. In Sec.~\ref{sec:formalism} 
we give an overview of eccentric orbits on a Schwarzschild background
and our FD method for solving the first-order field equations sourced by
such orbits. In Sec.~\ref{sec:PN} we demonstrate our method for
finding analytic expressions for the retarded MP in a double 
PN/small-eccentricity expansion. 
Sec.~\ref{sec:UAvgBackground} discusses the details of the 
gauge invariant we compute.
Sec.~\ref{sec:results} gives results
for both the MP itself, as well as $\langle U \rangle$, showing the 
merits of our re-summation of the small-$e$ series. We finish with a brief 
discussion in Sec.~\ref{sec:outlook}. Our Appendix provides details on
our low-order modes
and $\langle U \rangle$ written in a 
slightly different form, for comparison purposes.

Throughout, we use the $(-,+,+,+)$ metric signature and set $c = G = 1$,
(although we briefly use $\eta = c^{-1}$ for PN power counting).
Lowercase Greek indices run over Schwarzschild coordinates, $t,r,\th, \vp$.
We make use of the Martel and Poisson \cite{MartPois05} $\mathcal{M}^2 \times \mathcal{S}^2$
decomposition. Following their notation, lowercase Latin indices indicate $t$ or $r$
while uppercase Latin indices are either $\th$ or $\vp$.

\section{Inhomogeneous solutions to the first-order field equations in RWZ gauge}
\label{sec:formalism}
In this section we cover our method for solving the RWZ-gauge first-order
field equations in the FD, taking the source to be a point particle in eccentric
orbit. With the exception of a brief discussion of the singular structure of 
RWZ gauge at the end of this section, we follow closely the more detailed 
presentation of Ref.~\cite{HoppEvan10}.
We will work in Schwarzschild coordinates where the metric 
takes the standard form,
\begin{align}
ds^2 = -f dt^2 + f^{-1} dr^2 + r^2 d \th^2 + r^2 \sin^2 \th d \vp^2,
\end{align}
with $f \equiv 1 -2M/r$.

\subsection{Bound orbits on a Schwarzschild background}
\label{sec:orbits}

Let a small body, or particle, of mass $\mu$ orbit a static 
black hole of mass $M$, assuming
$q \equiv \mu / M \ll 1$. We parametrize the particle's background geodesic by 
proper time $\tau$, writing 
$x_p^{\a}(\tau) =\left[t_p(\tau),r_p(\tau), \pi/2, \varphi_p(\tau)\right]$.
Here and subsequently we use the subscript $p$ to indicate a quantity evaluated
on the worldline. Note that by choosing $\th_p = \pi / 2$ we have confined the 
particle to the equatorial plane with no loss of generality. 
Differentiating $x_p$ yields the four velocity 
\be
\label{eqn:four_velocity}
u^\a = \l \frac{{\mathcal{E}}}{f_{p}}, u^r, 0, \frac{{\mathcal{L}}}{r_p^2} \r,
\ee
where we have defined the two constants of motion, the specific energy 
$\mathcal{E}$ and specific angular momentum $\mathcal{L}$.
The constraint $u^\a u_\a = -1$ implies the following relation,
\be
\label{eq:removeRDot}
\dot r_p^2(t) = f_{p}^2 - \frac{f_{p}^2}{{\cal{E}}^2} U^2_{p},
\q \q
U^2(r,{\cal{L}}^2) \equiv f \l 1 + \frac{{\cal{L}}^2}{r^2} \r,
\ee
where a dot indicates a coordinate time derivative.

Any geodesic on a Schwarzschild background can be parametrized using
$\mathcal{E}$ and $\mathcal{L}$. For bound, eccentric motion, however, it is
convenient to instead use
the (dimensionless) semi-latus rectum $p$ and the 
eccentricity $e$ (see \cite{CutlKennPois94,BaraSago10}).  
These two pairs of parameters are related by 
\be
\label{eqn:defeandp}
{\mathcal{E}}^2 = \frac{(p-2)^2-4e^2}{p(p-3-e^2)},
\q \q
{\mathcal{L}}^2 = \frac{p^2 M^2}{p-3-e^2} .
\ee
Bound orbits satisfy the inequality $p > 6 + 2 e$ \cite{CutlKennPois94}.

Using the $p,e$ parametrization, the radial position of the particle is
given as a function of Darwin's \cite{Darw59} relativistic anomaly $\chi$,
\be
r_p \l \chi \r = \frac{pM}{1+ e \cos \chi} .
\ee
As $\chi$ runs from $0 \to 2\pi$, the particle travels
one radial libration, starting at periapsis.  
The quantities $\tau_p$, $t_p$, and $\vp_p$ are found by solving
first-order differential equations in $\chi$,
\begin{align}
\begin{split}
\label{eqn:darwinEqns}
\frac{dt_p}{d \chi} &= 
 \frac{p^2 M}{(p - 2 - 2 e \cos \chi) (1 + e \cos \chi)^2}
\left[ \frac{(p-2)^2 - 4 e^2}{p - 6 - 2 e \cos \chi} \right]^{1/2},
\\
\frac{d \varphi_p}{d\chi} 
&= \left[\frac{p}{p - 6 - 2 e \cos \chi}\right]^{1/2} ,
\\
\frac{d\tau_p}{d \chi} &= \frac{M p^{3/2}}{(1 + e \cos \chi)^2} 
\left[ \frac{p - 3 - e^2}{p - 6 - 2 e \cos \chi} \right]^{1/2} .
\end{split}
\end{align}
There is an analytic solution for $\vp_p$,
\be
\vp_p(\chi) = \left(\frac{4 p}{p - 6 - 2 e}\right)^{1/2} \, 
F\left(\frac{\chi}{2} \, \middle| \, -\frac{4 e}{p - 6 - 2 e}  \right) ,
\ee
where $F(x|m)$ is the incomplete elliptic integral of the first kind 
\cite{GradETC07}. Note that there also exists a more extensive semi-analytic 
solution for $t_p$  \cite{FujiHiki09}, though we do not provide it here.

Eccentric orbits have two fundamental frequencies. The libration between
periapsis and apoapsis is described by
\be
\label{eqn:O_r}
\O_r \equiv   \frac{2 \pi}{T_r},
\q \q
T_r \equiv \int_{0}^{2 \pi} \l \frac{dt_p}{d\chi} \r d \chi.
\ee
Meanwhile, the average rate of azimuthal advance over one radial period is
\be
\label{eqn:O_phi}
\O_\varphi \equiv \frac{\vp_p(2 \pi)}{T_r} 
= \frac{4}{T_r} \left(\frac{p}{p - 6 - 2 e}\right)^{1/2} \, 
\mathcal{K}\left(-\frac{4 e}{p - 6 - 2 e}  \right) ,
\ee
with $\mathcal{K}(m)$ being the complete elliptic integral of the first kind
\cite{GradETC07}.  The two frequencies $\Omega_r$ and $\Omega_{\vp}$
are only equal in the Newtonian limit.

\subsection{Solutions to the time domain master equation}

In RWZ gauge the field equations for the MP amplitudes can be
reduced to a single wave equation for each $\ell m$ mode.
The equation is satisfied by a parity-dependent master function from which
the MP amplitudes can be readily recovered. When $\ell+m$ is odd, we use
the Cunningham-Price-Moncrief (CPM) function, $\Psi^o_{\ell m}$,
and when $\ell+m$ is even, we use the Zerilli-Moncrief (ZM) function, 
$\Psi^e_{\ell m}$. In the remainder of this subsection we will use $\Psi_{\ell m}$
with no superscript to refer to either the ZM or CPM variable.
In each case the master equation has the form
\be
\label{eqn:masterEqTD}
\left[-\frac{\pa^2}{\pa t^2} + \frac{\pa^2}{\pa r_*^2} - V_\ell (r) \right]
\Psi_{\ell m}(t,r) = S_{\ell m}(t,r),
\ee
where both the potential $V_\ell$ and the source term $S_{\ell m}$ are parity-dependent.
The variable $r_* = r + 2M \log (r/2M - 1)$ is the standard tortoise coordinate.
The source term $S_{\ell m}$ is of the form
\begin{align}
\label{eqn:sourceTD}
S_{\ell m} = G_{\ell m}(t) \, \delta[r - r_p(t)] + F_{\ell m}(t) \,
\delta'[r - r_p(t)],
\end{align}
where the coefficients $G_{\ell m}$ are $F_{\ell m}$ are smooth functions.
While it is certainly possible to solve Eqn.~\eqref{eqn:masterEqTD} directly 
in the TD, at present we are interested in a FD approach.
As such, we decompose both $\Psi_{\ell m}$ and $S_{\ell m}$ in Fourier series as
\begin{align}
\label{eqn:psiAndSSeries}
\Psi_{\ell m}(t,r) = \sum_{n=-\infty}^\infty X_{\ell mn}(r) \, e^{-i \o t} , \q \q
S_{\ell m}(t,r) = \sum_{n=-\infty}^\infty Z_{\ell mn}(r) \, e^{-i \o t}.
\end{align}
In these expressions, the frequency $\o \equiv \omega_{mn} = m\Omega_\vp + n\Omega_r$, 
which follows from the  bi-periodic source. Note that we use the notation
$X_{\ell mn} \equiv X_{\ell m \o_{mn}}$.
The Fourier series coefficients are found as usual by,
\begin{align}
X_{\ell mn}(r) = \frac{1}{T_r} \int_0^{T_r} dt \ \Psi_{\ell m}(t,r) 
\, e^{i \o t},
\q \q 
Z_{\ell mn}(r) = \frac{1}{T_r} \int_0^{T_r} dt \ S_{\ell m}(t,r) 
\, e^{i \o t} .
\label{eqn:FourierCoeffs}
\end{align}
Combining Eqns.~\eqref{eqn:masterEqTD} and \eqref{eqn:psiAndSSeries}
yields the FD master equation,
\be
\label{eqn:masterInhomogFD}
\left[\frac{d^2}{dr_*^2} +\omega^2 -V_\ell (r) \right]
X_{\ell mn}(r) = Z_{\ell mn} (r).
\ee
This equation has two causal homogeneous solutions. At spatial infinity
the ``up'' solution $\hat{X}^+_{\ell mn}$ trends to $e^{i \o r_*}$.
As $r_* \to - \infty$ at the horizon, the ``in'' solution $\hat{X}^-_{\ell mn}$
trends to $e^{-i \o r_*}$. Here we use a hat to emphasize that these 
are unnormalized homogeneous solutions. 

With a causal pair of linearly independent solutions, 
one would normally find the particular 
solution to Eqn.~\eqref{eqn:masterInhomogFD} through the method
of variation of parameters. However, in this case, 
the singular source \eqref{eqn:sourceTD}
leads to a Gibbs phenomenon which
spoils the exponential convergence when forming  
$\Psi_{\ell m}$ in Eqn.~\eqref{eqn:psiAndSSeries}. It is now standard
to use the  method of extended homogeneous solutions (EHS) \cite{BaraOriSago08,HoppEvan10}
to obtain exponential convergence of $\Psi_{\ell m}$ and all its derivatives
at all locations, including the particle's. The method is covered extensively
elsewhere, and so we simply recount the procedure here.

We start by performing a convolution integral between the homogeneous solutions
and the FD source, which yields normalization coefficients,
\be
\label{eqn:normC}
C_{\ell mn}^{\pm} 
= \frac{1}{W_{\ell mn}} \int_{r_{\rm min}}^{r_{\rm max}} dr
 \ \frac{\hat X^{\mp}_{\ell mn} (r) Z_{\ell mn} (r)}{ f(r)} ,
\ee
where $W_{\ell mn}$ is the (constant in $r$) Wronskian
\be
W_{\ell mn} = f(r) \left( \hat{X}^-_{\ell mn} \frac{d \hat{X}^+_{\ell mn}}{dr}
- \hat{X}^+_{\ell mn} \frac{d \hat{X}^-_{\ell mn}}{dr} \right) .
\ee
Note that the integral \eqref{eqn:normC} is formally over all $r$, but we write it here
as limited to the  libration range $r_{\rm min} \le r \le r_{\rm max}$, since
outside that region $Z_{\ell mn} = 0$.
We next form the FD EHS, 
\be
\label{eqn:FD_EHS}
X^\pm_{\ell mn} (r) \equiv C^{\pm}_{\ell mn} \hat X_{\ell mn}^\pm (r), \q \q r > 2M,
\ee 
and subsequently define the TD EHS,
\be
\label{eqn:TD_EHS}
\Psi^\pm_{\ell m} (t,r) 
\equiv \sum_n X^\pm_{\ell mn} (r) \, e^{-i \o t}, \q \q r > 2M.
\ee
The TD EHS are formed from a set of smooth functions and therefore
the sum \eqref{eqn:TD_EHS} is exponentially convergent for all $r > 2M$, and all $t$.
Finally, the particular solution to Eqn.~\eqref{eqn:masterEqTD} is of the weak form
\begin{align}
\Psi_{\ell m} (t,r) = \Psi^{+}_{\ell m}(t,r) \theta \left[ r - r_p(t) \right] +
\Psi^{-}_{\ell m}(t,r) \theta \left[ r_p(t) - r \right],
\end{align}
where $\th$ is the Heaviside distribution.

\subsection{Metric perturbation reconstruction}

The even-parity MP amplitudes are reconstructed from the ZM master function
via the relations
\begin{align}
\label{eq:ReconstructEven}
\begin{split}
K^{\ell m,\pm} (t,r) & =  f \pa_r \Psi^{e,\pm}_{\ell m} + A \, \Psi_{\ell m}^{e,\pm} , \\
h_{rr}^{\ell m,\pm} (t,r) &= \frac{\La}{f^2}
\left[ \frac{\la+1}{r} \Psi_{\ell m}^{e,\pm} - K^{\ell m,\pm} \right] 
+ \frac{r}{f} \pa_r K^{\ell m,\pm}, \\
h_{tr}^{\ell m,\pm} (t,r) &=   
r \pa_t \pa_r \Psi_{\ell m}^{e,\pm} 
+ r B  \pa_t \Psi_{\ell m}^{e,\pm} , \\
h_{tt}^{\ell m,\pm} (t,r) &= f^2 h_{rr}^{\ell m,\pm},
\end{split}
\end{align}
where $\La(r) \equiv \la + 3M / r$, $\la \equiv \l \ell+2 \r \l \ell-1 \r / 2$, and
\begin{align}
A(r) \equiv \frac{1}{r \La} 
\left[ \la(\la+1) + \frac{3M}{r} \l \la + \frac{2M}{r} \r \right], 
\q \q
B(r) \equiv \frac{1}{r f \La} 
\left[ \la \l 1 - \frac{3M}{r} \r - \frac{3M^2}{r^2}  \right].
\end{align}
The odd-parity MP amplitudes are be reconstructed
from the CPM variable,
\begin{align}
\label{eq:ReconstructOdd}
h_t^{\ell m,\pm} (t,r) = \frac{f}{2} \pa_r \l r \Psi_{\ell m}^{o,\pm} \r,
\q \q
h_r^{\ell m,\pm} (t,r) = 
\frac{r}{2 f} \pa_t \Psi_{\ell m}^{o,\pm} .
\end{align}
In these expressions we have included $\pm$ superscripts
to indicate that MP amplitudes can be reconstructed on either the
left or right side of the particle.

At last, the retarded MP (which we write as $p_{\mu \nu}$)
can be synthesized by multiplying by spherical harmonics and
summing over $\ell m$ modes. 
Our particular harmonic decomposition is due to Martel and Poisson \cite{MartPois05}.
The spherical harmonics used below (even-parity
scalar $Y^{\ell m}$ and odd-parity vector $X^{\ell m}_B$), 
along with the two-sphere metric ($\O_{AB}$), can be found in that reference.
Lowercase Latin indices run over $t$ and $r$ while 
uppercase Latin indices run over $\th$ and $\vp$.
To emphasize a point about
the singular nature of RWZ gauge, we perform the sum over spherical 
harmonics in two stages.
First, we form the $\ell m$ contribution to each MP component
(see Ref.~\cite{HoppEvan10}),
\begin{align}
\label{eqn:MPlm}
\begin{split}
p_{ab}^{\ell m} (x^\mu) &=  
\Big[ h_{ab}^{\ell m,+} (t,r) \th (z)
+ h_{ab}^{\ell m,-} (t,r) \th (-z)
+ h_{ab}^{\ell m,S} (t) \d ( z )
 \Big] Y^{\ell m} (\th, \vp) ,\\
 p_{aB}^{\ell m} (x^\mu) &= \Big[ h_a^{\ell m,+}(t,r) \th(z)
 +h_a^{\ell m,-}(t,r) \th(-z) \Big] X_B^{\ell m} (\th, \vp), \\
 p_{AB}^{\ell m} (x^\mu) &= \Big[ K^{\ell m,+} (t,r) \th (z) 
 + K^{\ell m,-} (t,r) \th (-z)  \Big] r^2 \O_{AB} Y^{\ell m} (\th, \vp).
 \end{split}
\end{align}
These expressions are written as weak solutions in terms of the 
Heaviside and Dirac distributions $\th$ and $\d$, which depend
on $z \equiv r - r_p(t)$.
Note that in general the Martel-Poisson decomposition also includes
scalar amplitudes $j^{\ell m}_a,G^{\ell m}$, and $h_2^{\ell m}$,
but these are set to zero in RWZ gauge.

The expressions in Eqn.~\eqref{eqn:MPlm} suggest that the
RWZ gauge retarded MP,
is not only discontinuous at the location of the particle,
but actually proportional to the Dirac delta function for certain components. 
Moreover, this singularity is spread over a two-sphere of radius $r = r_p(t)$.
However, we find that, at least at the particle's location, 
this is in fact not true. Indeed, setting $\th = \th_p(t) = \pi/2$, $\vp = \vp_p (t)$ 
and summing the expressions
\eqref{eqn:MPlm} over $m$ exactly cancels out the delta functions, and the 
remaining amplitudes are actually continuous at $r=r_p$ for all $\ell \ge 2$.
Thus, we write
\begin{align}
\begin{split}
p_{ab}^{\ell} (x^\mu_p) & = 
\sum_m  h_{ab}^{\ell m,\pm} (t_p,r_p) Y^{\ell m} (\pi/2, \vp_p) ,\\
p_{aB}^{\ell} (x^\mu_p) & = 
\sum_m  h_{a}^{\ell m,\pm} (t_p,r_p) X_B^{\ell m} (\pi/2, \vp_p) ,\\
p_{AB}^{\ell} (x^\mu_p) & = 
r_p^2 \ \O_{AB} \sum_m  K^{\ell m,\pm} (t_p,r_p) Y^{\ell m} (\pi/2, \vp_p),
\label{eqn:MPAtParticleL}
\end{split}
\end{align}
and a single-valued RHS becomes a check on all calculations. The full retarded
MP is then a simple sum over all $\ell$,
\begin{align}
p_{ab} (x^\mu_p) = \sum_{\ell=0}^\infty p_{ab}^{\ell} (x^\mu_p), \q \q
p_{aB} (x^\mu_p) = \sum_{\ell=0}^\infty p_{aB}^{\ell} (x^\mu_p), \q \q
p_{AB} (x^\mu_p) = \sum_{\ell=0}^\infty p_{AB}^{\ell} (x^\mu_p).
\end{align}
Zerilli's analytic $\ell=0,1$ solutions are given in App.~\ref{sec:lowOrder}.

\section{Post-Newtonian solutions}
\label{sec:PN}

We now present our method for finding the retarded MP induced by a particle in 
eccentric motion about a Schwarzschild black hole. 
We combine the formalism of the previous section with PN expansions 
of all relevant quantities. Our final results, given in Sec.~\ref{sec:results},
include expansions to 4PN (that is, 4 terms beyond leading order). 
At each PN order we expand to 10th order in eccentricity.
For the purposes of pedagogy and space, however, in this section
we keep only two PN orders and (where necessary) two terms in
the small-$e$ expansion. 
Our final solutions will give the MP at the location
of the particle for each $\ell$-mode, as in Eqn.~\eqref{eqn:MPAtParticleL}. 
Note that the presentation
from Sec.~\ref{sec:pnHomog} onward is only relevant to modes $\ell \ge 2$.

\subsection{Post-Newtonian expansions of orbit quantities}
\label{sec:pnOrb}

\subsubsection{Position-independent orbit quantities}

We take as our PN expansion parameter the inverse of the 
dimensionless semi-latus rectum $p$.
Assuming it to be small, we expand $dt_p/d\chi$ from Eqn.~\eqref{eqn:darwinEqns}.
Inserting the resulting expansion in Eqn.~\eqref{eqn:O_r}
we integrate order-by-order and find
\begin{align}
\O_r = \frac{1}{M} \l \frac{1-e^2}{p} \r^{3/2} 
\left[ 1 - 3 \frac{1-e^2}{p} + \mathcal{O} \l p^{-2} \r \right] .
\end{align}
Expanding Eqn.~\eqref{eqn:O_phi} in the same limit gives
\begin{align}
\label{eqn:O_phiPN}
\O_\vp = \frac{1}{M} \l \frac{1-e^2}{p} \r^{3/2} 
\left[ 1 + 3 \frac{e^2}{p} + \mathcal{O} \l p^{-2} \r \right] .
\end{align}
As expected $\O_r = \O_\vp$ at Newtonian order.
We now introduce the dimensionless gauge invariant PN parameter $y$,
\begin{align}
 y \equiv \left( M \Omega_\vp \right)^{2/3}.
\label{eqn:yPN}
\end{align} 
Combining Eqns.~\eqref{eqn:O_phiPN}
and ~\eqref{eqn:yPN} we find a PN expansion for $y$ in terms of $p$.
We invert the expansion to get $p$ in terms of $y$
\begin{align}
p = \frac{1-e^2}{y} + 2 e^2 + \mathcal{O} \l y^1 \r.
\end{align}
This allows us to obtain expansions for both $\Omega_r$ and 
$\Omega_\vp$ in powers of $y$,
\begin{align}
\Omega_r &= \frac{y^{3/2}}{M} 
\left[ 1-\frac{3}{1-e^2}y+\mathcal{O}(y^2) \right],\q \q
\Omega_\vp = \frac{y^{3/2}}{M}.
\end{align}
Note that the $\Omega_\vp$ expression is exact due to the definition of $y$.
With the above expressions, we are able to find
PN expansions of all orbit quantities in terms of $y$. 
The specific energy and angular momentum 
follow from Eqn.~\eqref{eqn:defeandp},
\begin{align}
\mathcal{E} &= 1-\frac{y}{2}
+\frac{3 + 5 e^2}{8-8 e^2}y^2
+\mathcal{O}\left(y^3\right),
\q \q
\mathcal{L} = 
\sqrt{1-e^2} M y^{-1/2}
+\frac{3 M \left(1+e^2\right)}{2 \sqrt{1-e^2}} y^{1/2}
+\mathcal{O}\left(y^{3/2}\right).
\end{align}
Later, we will also need the radial period as measured in coordinate time
[Eqn.~\eqref{eqn:O_r}] and proper time
(found by integrating $d \tau_p / d\chi$ from $0-2 \pi$). They are,
respectively
\begin{align}
T_r = \frac{2 \pi M}{y^{3/2}}
\left[1+\frac{3}{1-e^2} y +\mathcal{O}\left(y^{2}\right) \right], 
\q \q
\mathcal{T}_r = \frac{2 \pi M}{y^{3/2}}
\left[ 1+\frac{3 + 3 e^2}{2-2 e^2} y+\mathcal{O}\left(y^{2}\right) \right].
\end{align}

\subsubsection{Position-dependent orbit quantities}

We now expand quantities that vary along the worldline as functions of the
relativistic anomaly $\chi$.
We start with the radial position 
$r_p$, first expanding in the PN parameter $y$, and then at each PN order
we expand in eccentricity $e$. The resulting double expansion (here keeping only
the first two non-zero orders in each $y$ and $e$) is
\begin{align}
\label{eqn:rpPN}
r_p (\chi) &= 
M \Big[ 1- e \cos \chi +\mathcal{O}\left(e^2\right) \Big] y^{-1} 
+
M \Big[ 2 e^2-2 e^3 \cos \chi+\mathcal{O}\left(e^4\right)\Big] y^0
+
\mathcal{O}\left(y^1\right).
\end{align}
Next, consider the $\vp_p(t)$ 
motion, which can be decomposed into two parts as \cite{BaraOriSago08}
\be
\label{eqn:phiSplit}
\varphi_p(t) = \Omega_{\varphi} t + \Delta\varphi(t).
\ee
The first term represents the mean azimuthal advance, while the second
term is periodic in $T_r$. In our expansions we avoid ever using $\Omega_\vp t$
explicitly, and work only with the $\Delta \vp$. This is convenient because 
terms involving $\Omega_\vp t$ can lead linear-in-$\chi$ terms. For example,
notice that after leading order in $y$, 
$e^{i m \Omega_\vp t}$ is not strictly oscillatory,
\begin{align}
\begin{split}
e^{i m \Omega_\vp t}
&=
\Big[
e^{i m \chi}-2 i m  \sin (\chi) e^{i m \chi}  e  +\mathcal{O}\left(e^2\right)
\Big] y^0 \\
& \hspace{25ex}
 +
\Big[3 i m \chi e^{i m \chi} +3 m (2 m \chi -i) \sin (\chi) e^{i m \chi} e
+\mathcal{O}\left(e^2\right)\Big] y
+
\mathcal{O}\left(y^{2}\right) .
\end{split}
\end{align}
On the other hand, $e^{i n \Omega_r t}$ \emph{is} oscillatory for all PN orders,
\begin{align}
e^{i n \Omega_r t}
&=
\Big[ e^{i n \chi }-2 i n \sin (\chi) e^{i n \chi} e +\mathcal{O}\left(e^2\right)\Big] y^0
+
\Big[3 i n \sin (\chi) e^{i n \chi} e+\mathcal{O}\left(e^2\right)\Big] y
+
\mathcal{O}\left(y^{2}\right).
\end{align}
This qualitative difference between the fundamental frequencies
can be traced back to the fact that $\Omega_\vp$ is a ``rotation-type''
frequency describing average accumulation of phase while
$\Omega_r$ is a ``libration-type'' frequency describing periodic motion in $r$.
Since $\Delta \vp$ is periodic in $T_r$ we find that its expansion is free of linear-in-$\chi$
terms,
\begin{align}
\Delta \vp (\chi) &=
\Big[2 e \sin \chi  +\mathcal{O}\left(e^2\right)\Big] y^0
+\Big[4 e \sin \chi +\mathcal{O}\left(e^2\right)\Big] y
+\mathcal{O}\left(y^{2}\right).
\end{align}
Lastly, because they will be useful later, we also note the expansions
\begin{align}
\begin{split}
e^{-i m \Delta \vp(\chi)} &= 
\Big[1-2 i m e \sin \chi +\mathcal{O}\left(e^2\right) \Big] y^0
+
\Big[ - 4 i m e \sin \chi+\mathcal{O}\left(e^2\right)\Big] y
+
\mathcal{O}\left(y^{2}\right),\\
\frac{dt_p}{d\chi}
&=
M \Big[ 1-2  e \cos \chi+\mathcal{O}\left(e^2\right) \Big] y^{-3/2}
+
M \Big[ 3 -3   e \cos \chi+\mathcal{O}\left(e^2\right) \Big] y^{-1/2}
+
\mathcal{O}\left(y^{1/2}\right), \\
\frac{d\tau_p}{d\chi}
&=
M\Big[1 -2   e \cos \chi+\mathcal{O}\left(e^2\right) \Big] y^{-3/2}
+
M \left[ \frac{3}{2}-2   e \cos \chi+\mathcal{O}\left(e^2\right) \right] y^{-1/2}
+
\mathcal{O}\left(y^{1/2}\right).\\
\end{split}
\end{align}

\subsection{Frequency domain homogeneous solutions}
\label{sec:pnHomog}

We now derive expressions for the unnormalized FD master function evaluated, at the
particle's location. We start with homogeneous solutions 
to the odd-parity master equation of the form derived in
Refs.~\cite{BiniDamo13,KavaOtteWard15}. These solutions are 
written as a double expansion in
small frequency and large $r$ using the dimensionless expansion parameters
\begin{align}
\mathcal{X}_1 \equiv \frac{M}{r},
\quad \quad 
\mathcal{X}_2 \equiv (\omega r)^2,
\label{eqn:Xdefs}
\end{align}
which are assumed to be the same order of magnitude.
Note that we choose the symbols $\mathcal{X}_1$ and $\mathcal{X}_2$
rather than the standard  $X_1$ and $X_2$ to avoid confusion with the FD 
master function.
Written in terms of these variables, the $\ell = 2$
homogeneous solutions to the odd-parity master equation are
\begin{align}
\label{eqn:xOdd2}
\hat{X}_2^{o,+}
&=
\mathcal{X}_1^2 \eta^4
+\frac{1}{6} \mathcal{X}_1^2 \left(10 \mathcal{X}_1 +\mathcal{X}_2\right)\eta ^6 
+
\mathcal{O}\left(\eta ^8\right),
\q \q
\hat{X}_2^{o,-}
=
\frac{1}{\mathcal{X}_1^3} \eta^{-6}
-\frac{\mathcal{X}_2}{14 \mathcal{X}_1^3} \eta^{-4}
+
\mathcal{O}\left(\eta^{-2}\right),
\end{align}
with $+$ indicating the infinity-side (up) solution, and $-$ the 
horizon-side (in) solution. 
We use $\eta = c^{-1}$ to keep track of the PN order.
The even-parity solutions are computed from the 
Chandrasekhar-Detweiler 
transformation \cite{Chan75}
\begin{align}
\label{eqn:chandra}
\hat{X}^e_\ell
&= 
\left[
\frac{(\ell-1) \ell (\ell+1) (\ell+2)}{24} 
+
\frac{3 (1-2 \mathcal{X}_1) \mathcal{X}_1^2}{(\ell-1) (\ell+2)+6 \mathcal{X}_1}
\right] 
\hat{X}^o_\ell
+
 \frac{M (1-2 \mathcal{X}_1)}{2}\frac{d\hat{X}^o_\ell}{dr},
\end{align}
which is true to all PN orders.
We compute $r$ derivatives with the relation
\begin{align}
\frac{d}{dr} = 
\frac{\mathcal{X}_1}{M}  \l \sqrt{\mathcal{X}_2} \frac{d}{d \sqrt{\mathcal{X}_2}}
 - 
\mathcal{X}_1  \frac{d}{d \mathcal{X}_1} \r.
\end{align}
Then, the even-parity $\ell=2$ homogeneous solutions are 
\begin{align}
\begin{split}
\hat{X}_2^{e,+}
&=
\mathcal{X}_1^2 \eta^4 
+
\left(
\frac{2 \mathcal{X}_1^3}{3} + \frac{\mathcal{X}_1^2 \mathcal{X}_2}{6}
\right)\eta^6 
+\mathcal{O}\left(\eta ^8\right)
,	\q \q 
\hat{X}_2^{e,-}
=
\frac{1}{\mathcal{X}_1^3} \eta^{-6} 
+ 
\l \frac{3}{2 \mathcal{X}_1^2}-\frac{\mathcal{X}_2}{14 \mathcal{X}_1^3} \r \eta^{-4}
+
\mathcal{O}\left(\eta^{-2}\right).
\end{split}
\end{align}
For the remainder of this presentation we focus on the
specific example of infinity-side, odd-parity.
The even-parity and horizon-side calculations, as well as those
for the $r$-derivatives of the master function, follow from
an equivalent procedure.

We are interested in evaluating the homogeneous solutions
at the location of the particle.
Plugging in Eqn.~\eqref{eqn:Xdefs} with $r=r_p(t)$ we find
\begin{align}
\hat{X}_2^{o,+}
&=
\frac{M^2}{r_p^2}\eta^4
+
\frac{M^2 \left(10 M+\omega ^2 r_p^3\right)}{6 r_p^3} \eta^6
+
\mathcal{O}\left(\eta^8\right).
\label{eqn:homogOddROm}
\end{align}
In this expression $\omega$ and $r_p$ are valid to all PN orders.
To make further progress, we now expand $\omega$ 
and $r_p$ in both the PN parameter $y$ 
and the eccentricity $e$. The double expansion of $\omega$
(again keeping only the first two non-zero orders in $y$ and $e$) is
\begin{align}
\o = \omega_{mn} &= 
\frac{ (m+n)}{M}y^{3/2}
+ 
\left[-\frac{3 n}{M}-\frac{3 n e^2}{M} +\mathcal{O}\left(e^3\right)\right] y^{5/2}
+
\mathcal{O}\left(y^{7/2}\right).
\end{align}
Inserting this, along with the PN expression for $r_p$
from Eqn.~\eqref{eqn:rpPN} into Eqn.~\eqref{eqn:homogOddROm} we find 
(dropping $\eta$ in favor
of $y$ for counting PN orders)
\begin{align}
\hat{X}^{o,+}_{2mn} &=
\Big[ 1+2 e \cos \chi + \mathcal{O}\left(e^2\right) \Big] y^2 
+
\left[
\frac{1}{6} \left(m^2+2 n m+n^2+10\right)
+
5 e \cos \chi
+\mathcal{O}\left(e^2\right)\right]y^3 
+
\mathcal{O}\left(y^4\right)	.
\label{eqn:homogOddXE}
\end{align}
This expression, while a function of $\chi$, which tracks the particle, is
still just a homogeneous solution to the FD master equation.
Note that Eqn.~\eqref{eqn:homogOddXE} is specific to $\ell=2$, 
but is valid 
for $m = \pm 1$ and any $n$ (though we will see that a finite $e$ expansion
limits the number of relevant harmonics $n$).

\subsection{Frequency domain extended homogeneous solutions}

Our next step is to find normalization coefficients so that we can
form the FD EHS, as in Eqn.~\eqref{eqn:FD_EHS}.
In Eqn.~\eqref{eqn:normC} we wrote an expression for $C^\pm_{\ell m n}$
that depends on a generic source bounded between $r_{\rm min}$ and $r_{\rm max}$.
We now specify to the form given in Eqn.~\eqref{eqn:sourceTD}.
It can be shown \cite{HoppEvan10} that such a source, when combined with 
Eqn.~\eqref{eqn:FourierCoeffs}, leads to a normalization 
integral of the form
\begin{align}
\label{eqn:EHSC}	
C_{\ell mn}^\pm  
&=   \frac{1}{W_{\ell mn} T_r} \int_0^{T_r}
\Bigg[ 
 \frac{1}{f_{p}} \hat X^\mp_{\ell mn}
 G_{\ell m} + \l \frac{2M}{r_{p}^2 f_{p}^{2}} \hat X^\mp_{\ell mn}
 - \frac{1}{f_{p}} 
 \frac{d \hat X^\mp_{\ell mn}}{dr} \r F_{\ell m}
 \Bigg]  e^{i \o t}  \, dt, 
\end{align}
The terms $G_{\ell m}$ and $F_{\ell m}$ arise from taking spherical harmonic projections 
of the point particle's stress energy tensor, and
as such, each carries a factor of $e^{- im \vp_p(t)}$. 
Noting Eqn.~\eqref{eqn:phiSplit}, 
it is useful to remove the
$e^{- im \Omega_{\varphi} t}$ contribution from $G_{\ell m}$ and $F_{\ell m}$
by defining \cite{HoppETC15}
\begin{align}
\label{eqn:GFBar}
\bar{G}_{\ell m}(t) &\equiv G_{\ell m}(t) \, e^{im\Omega_{\vp} t}, 
\quad \quad
\bar{F}_{\ell m}(t) \equiv F_{\ell m}(t) \, e^{im\Omega_{\vp} t},
\end{align}
which are $T_r$-periodic.
The $e^{- im \Omega_{\varphi} t}$ term cancels with a compensating term in the
$e^{i \o t}$ of Eqn.~\eqref{eqn:EHSC}. Thus, changing the integration
variable to $\chi$, we are left with
\begin{align}
C_{\ell mn}^\pm  
&=   \frac{1}{W_{\ell mn} T_r} \int_0^{2 \pi}
\Bigg[ 
 \frac{1}{f_{p}} \hat X^\mp_{\ell mn}
\bar{G}_{\ell m} + \l \frac{2M}{r_{p}^2 f_{p}^{2}} \hat X^\mp_{\ell mn}
 - \frac{1}{f_{p}} 
 \frac{d \hat X^\mp_{\ell mn}}{dr} \r \bar{F}_{\ell m}
 \Bigg]  e^{i n \O_r t}  \, \frac{dt_p}{d\chi} d\chi,
 \label{eqn:CFromGFBar}
\end{align}
and all terms in the integrand are now $2 \pi$-periodic (when considered as 
functions of $\chi$). 

Generally, we compute the integral \eqref{eqn:CFromGFBar}
numerically. In the PN/small-eccentricity regime, however, we are able to 
perform the integral analytically $\ell$-by-$\ell$. In previous subsections,
we have already computed PN expansions 
of $\hat{X}^\mp_{\ell mn}$, $r_p$, $dt_p/d\chi$, and  $e^{i n \O_r t}$.
The $f_p$ terms follow naturally from the $r_p$ expansion. What remains 
is the expansion of $\bar{G}_{\ell m}$ and $\bar{F}_{\ell m}$.

In the odd-parity sector we have 
\begin{align}
\begin{split}
G^o_{\ell m} &=
\frac{32 \pi  \mu  \mathcal{L} f_p}{(\ell-1) \ell (\ell+1) (\ell+2) \mathcal{E}^2 r_p^5}
\Big\{
\mathcal{L} \mathcal{E} r_p^2 \dot{r}_p X_{\vp \vp}^{\ell m,*}
-
f_p \Big[ 
5 M r_p^2+7 M \mathcal{L}^2+\left(2 \mathcal{E}^2-1\right) r_p^3 -2 \mathcal{L}^2 r_p
\Big] X_{\vp}^{\ell m,*}
 \Big\} ,  \\
F^o_{\ell m} &=
\frac{32 \pi  \mu  \mathcal{L} f_p^3 \left(r_p^2+\mathcal{L}^2\right)}{(\ell-1) \ell (\ell+1) (\ell+2) \mathcal{E}^2 r_p^3} X_{\vp}^{\ell m,*} .
\end{split}
\end{align}
The terms $X_{\vp}^{\ell m,*}$ and $X_{\vp \vp}^{\ell m,*}$ are complex conjugates
of odd-parity vector and tensor spherical harmonics, evaluated on the worldline. 
See Ref.~\cite{HoppEvan10} for details.
We now insert the various expansions we have already computed,
as usual only keeping two terms in each $y$ and $e$. The resulting $2\pi$-periodic 
expressions are
\begin{align}
\begin{split}
\bar{G}^o_{\ell m} 
&=
\frac{16 \pi \mu}{M} 
\frac{\pa_\th Y_{\ell m} (\pi/2,0)}{ (\ell-1) \ell (\ell+1) (\ell+2)}
\bigg\{ 
2 \Big[-1 - 2( \cos \chi - i m \sin \chi) e
+
\mathcal{O}\left(e^2\right)\Big] y^{3/2} 
\\
& \hspace{40 ex} +
\Big[ 1 + 4 (2 \cos \chi + i m \sin \chi) e
+
\mathcal{O}\left(e^2\right)\Big] y^{5/2} 
+
\mathcal{O}\left(y^{7/2}\right)
\bigg\},
\\
\bar{F}^o_{\ell m} 
&=
16 \pi \mu 
\frac{\pa_\th Y_{\ell m} (\pi/2,0)}{(\ell-1) \ell (\ell+1) (\ell+2)}
\bigg\{ 
2\Big[1 + (\cos \chi - 2 i m \sin \chi) e
+
\mathcal{O}\left(e^2\right)\Big] y^{1/2}
\\
& \hspace{40 ex} 
-
\Big[ 5+(13 \cos \chi-2 i m \sin \chi) e
+
\mathcal{O}\left(e^2\right)\Big] y^{3/2}
+
\mathcal{O}\left(y^{5/2}\right)
\bigg\}.
\end{split}
\end{align}

Having expanded all the relevant quantities, we are now in a position to perform the
integral \eqref{eqn:CFromGFBar} analytically order-by-order. The integral itself 
is straightforward; we encounter nothing more than complex exponentials.
The resulting normalization coefficient is
\begin{align}
C^{o,+}_{2mn} &=
\mu \ \pa_\th Y_{2 m}(\pi/2, 0) \sin (n \pi) 
\Bigg\{
\frac{16}{15}\left[\frac{1}{n}+\frac{(n-2 m)}{n^2-1}  e 
+\mathcal{O}\left(e^2\right) \right] y^{-3/2}\\
+ 
 \frac{4}{105} &
\left[
-\frac{\left(3 m^2+6 m n+3 n^2+14\right)}{n} 
+
\frac{\left(6 m^3+3 m^2 n-m \left(12 n^2+77\right)-9 n^3+49 n\right)}{n^2-1}  e
+\mathcal{O}\left(e^2\right)
\right] y^{-1/2}
+\mathcal{O}\left(y^{1/2}\right)
\Bigg\}. \notag
\end{align}
Note that there are several terms in this expression
which appear at first glance to diverge for integer values of $n$, 
but all values are finite in the limit.
With the normalization coefficients in hand, the FD EHS are just the product of 
$\hat{X}^{\pm}_{\ell mn}$ and $C^{\pm}_{\ell mn}$ as shown in Eqn.~\eqref{eqn:FD_EHS},
\begin{align}
\begin{split}
X^{o,+}_{2mn} &=
\mu \ \pa_\th Y_{2 m}(\pi/2, 0)  \sin (\pi  n) 
\Bigg\{
\frac{16}{15}
\left[ 
\frac{1}{n}
+
\left(\frac{n-2 m}{n^2-1} + \frac{2}{n} \cos \chi \right) e
+
 \mathcal{O}\l e^2\r
\right] y^{1/2}  \\
&
+
\frac{4}{105}
\Bigg[
\frac{5 m^2+10 m n+5 n^2+98}{3 n}
-
\Bigg(
\frac{2 \left(3 m^2+6 m n+3 n^2-56\right)}{n} \cos \chi \\
& \hspace{15ex}
+
\frac{10 m^3+33 m^2 n+36 m n^2+511 m+13 n^3-287 n}{3\left(n^2-1\right)}
\Bigg)
e+
\mathcal{O}\l e^2\r
\Bigg] y^{3/2} 
+  
\mathcal{O}\l y^{5/2}\r
\Bigg\}.
\end{split}
\end{align}

\subsection{Radially periodic, time domain extended homogeneous solutions}

We are now in a position to return to the TD by summing over harmonics $n$, 
as in Eqn.~\eqref{eqn:TD_EHS}. Since we have performed the $e$-expansion
to a finite order, the sum \eqref{eqn:TD_EHS} actually truncates. We find
that if we compute the FD EHS with eccentricity contributions up to $e^N$,
then the only non-zero terms in Eqn.~\eqref{eqn:TD_EHS} are 
in the range $-N \le n \le N$.

Another wrinkle comes into play at this stage. The full TD EHS are formed 
from a sum involving $e^{-i \o t} = e^{-i (m \O_\vp + n \O_r)t}$. 
As discussed in Sec.~\ref{sec:pnOrb}, however,
the factor $e^{-i m \O_\vp t}$ is not purely oscillatory when expanded in $y$ and $e$.
It is therefore more useful to form the quantity 

\smallskip

\be
\bar{\Psi}^\pm_{\ell m} (t,r) \equiv
\Psi^\pm_{\ell m} (t,r) e^{i m \Omega_\vp t}
= 
\sum_n X^\pm_{\ell mn} (r) \, e^{-i n \Omega_r t}.
\label{eqn:psiBar}
\ee
Contrary to $\Psi^\pm_{\ell m}$, $\bar{\Psi}^\pm_{\ell m}$ 
is purely periodic in $\chi$, with no linear terms.
This leads one to ask: don't we need that $e^{- i m \Omega_\vp t}$ term? 
Surprisingly, the answer is no, at least when computing local quantities.
The reason is that whenever we want to compute anything physically relevant, 
like the GSF, we have to
multiply by spherical harmonics and sum over the $\ell m$ modes. The spherical harmonics 
carry an exactly compensating term $e^{i m \Omega_\vp t}$ which cancels this piece out.

We have been considering the example of odd-parity, $\ell=2$ expanded to 
include two powers of $y$ and $e$. We can form $\bar{\Psi}_{2m}^{o,\pm}$ 
by summing over $n=-1,0,1$, and so we find
\begin{align}
\label{eqn:psiBarOdd}
\begin{split}
\bar{\Psi}_{2m}^{o,+} &=
\mu \pi \ \pa_\th Y_{2 m}(\pi/2, 0) 
\Bigg\{
\frac{16}{15}
\Big[
1+\big(\cos \chi - 2 i m \sin \chi \big) e+\mathcal{O}\left(e^2\right)
\Big]y^{1/2} \\
& \hspace{5ex} +
\frac{4}{315}\Big[ 
98 + 5 m^2
+
\Big(
   \left(15 m^2+62\right) \cos \chi 
- i m \left(10 m^2+547\right) \sin \chi \Big) e
+\mathcal{O}\left(e^2\right)\Big]
   y^{3/2} 
 +\mathcal{O}\left(y^{5/2}\right)
\Bigg\}.
\end{split}	
\end{align}
Note that we need only perform the calculation once for 
each $\ell$ mode. Therefore, this expression is valid for all
$-2 \le m \le 2$ (although in this case we are only interested in 
$m=\pm 1$, of course).

\subsection{Metric perturbation reconstruction}

We now wish to use the master function to reconstruct the retarded MP
at the location of the particle.
We first form the MP amplitudes, and then sum over spherical harmonics to
form the full MP.
Expressions for the MP amplitudes are given earlier in 
Eqns.~\eqref{eq:ReconstructEven} and \eqref{eq:ReconstructOdd}. However,
as with the master function, we wish to form``barred'' versions of the 
MP amplitudes, which are periodic in $T_r$,
e.g.~$\bar{h}^{\ell m}_{tt}=h^{\ell m}_{tt} e^{i m \Omega_\vp t}$.
Note that since the full master function contains an extra factor of
$e^{-i m \Omega_\vp t}$, time derivatives of $\Psi_{\ell m}$ pick up counter terms
when written in terms of $\bar{\Psi}_{\ell m}$. 
The adjusted reconstruction expressions 
are, for even parity
\begin{align}
\label{eq:ReconstructEvenBar}
\begin{split}
\bar{K}^{\ell m,\pm} (\chi) & =  f_p \pa_r \bar{\Psi}^{e,\pm}_{\ell m} (\chi)
  + A_p \, \bar{\Psi}_{\ell m}^{e,\pm} (\chi) , \\
\bar{h}_{rr}^{\ell m,\pm} (\chi) &= \frac{\La_p}{f_p^2}
\left[ \frac{\la+1}{r_p} \bar{\Psi}_{\ell m}^{e,\pm} (\chi) 
- \bar{K}^{\ell m,\pm} (\chi) \right] 
+ \frac{r_p}{f_p} \pa_r \bar{K}^{\ell m,\pm} (\chi), \\
\bar{h}_{tr}^{\ell m,\pm} (\chi) &=   
r_p \Big[ \pa_t \pa_r \bar{\Psi}_{\ell m}^{e,\pm} (\chi) 
- i m \O_\vp \pa_r \bar{\Psi}_{\ell m}^{e,\pm} (\chi) \Big]
+ r_p B_p \, \Big[ \pa_t \bar{\Psi}_{\ell m}^{e,\pm} (\chi)
 - i m \O_\vp \bar{\Psi}_{\ell m}^{e,\pm} (\chi) \Big], \\
\bar{h}_{tt}^{\ell m,\pm} (\chi) &= f_p^2 \bar{h}_{rr}^{\ell m,\pm} (\chi),
\end{split}
\end{align}
and odd parity
\begin{align}
\label{eq:ReconstructOddBar}
\bar{h}_t^{\ell m,\pm} (\chi) & =
\frac{f_p}{2} \Big[ \bar{\Psi}_{\ell m}^{o,\pm} (\chi)
+
r_p  \pa_r  \bar{\Psi}_{\ell m}^{o,\pm} (\chi) \Big],
\q \q
\bar{h}_r^{\ell m,\pm} (\chi) =
\frac{r_p}{2 f_p} \Big[ \pa_t \bar{\Psi}_{\ell m}^{o,\pm} (\chi) 
- i m \O_\vp \bar{\Psi}_{\ell m}^{o,\pm} (\chi) \Big].
\end{align}
Note the subscript $p$ on all $r$-dependent quantities indicating evaluation
at $r=r_p(\chi)$. Be aware that the functional-$\chi$ notation that we use
is shorthand for, e.g. 
$\bar{\Psi}_{\ell m}^{e,\pm} (\chi) = \bar{\Psi}_{\ell m}^{e,\pm} 
\left[t_p(\chi), r_p(\chi) \right]$. Therefore, it is \emph{not} true
that $\pa_t \bar{\Psi}_{\ell m}^{e,\pm} (\chi) = 
\pa_\chi \bar{\Psi}_{\ell m}^{e,\pm} (\chi) (dt_p/d\chi)^{-1}$.
To the contrary, all $t$ and $r$ derivatives of $\bar{\Psi}^{e/o, \pm}_{\ell m}$
must be computed explicitly. The $r$ derivatives follow from 
forming the FD EHS $\pa_r X_{\ell m n}^{e/o, \pm}$, while $t$ derivatives
are formed by taking the $t$ derivative of Eqn.~\eqref{eqn:psiBar}.

Next, we form the MP components for each $\ell m$ mode, as shown 
in Eqn.~\eqref{eqn:MPlm}. Now though, with the ``barred'' form of the MP
amplitudes, we must multiply by the ``barred'' form of the various 
spherical harmonics, e.g. 
$\bar{Y}^{\ell m} = Y^{\ell m} \left[ \pi/2 , \vp_p(t) \right] e^{- i m \O_\vp t}$.
This is not to be confused with complex conjugation, which we indicate with
an asterisk ($*$).
As expected, this does not change quantities evaluated on the worldline. For example
\begin{align}
p_{tt}^{\ell m} (\chi) = 
h_{tt}^{\ell m} Y^{\ell m}
= \bar{h}_{tt}^{\ell m} e^{i m \O_\vp t} e^{- i m \O_\vp t} \bar{Y}^{\ell m}	
= \bar{h}_{tt}^{\ell m} \bar{Y}^{\ell m}.
\end{align}
Thus, the MP components for each $\ell m$ mode can be
formed on either side of the particle through the expressions
\begin{align}
\label{eqn:mpBarlm}
\begin{split}
p_{ab}^{\ell m,\pm} (\chi) &=  \bar{h}_{ab}^{\ell m,\pm} (\chi) 
\bar{Y}^{\ell m,\pm}(\chi) , \\
 p_{aB}^{\ell m,\pm} (\chi) &= \bar{h}_a^{\ell m,\pm} (\chi) 
 \bar{X}_B^{\ell m,\pm} (\chi) , \\
 p_{AB}^{\ell m,\pm}  (\chi) &=  r_p^2  \O_{AB}
  \bar{K}^{\ell m,\pm} (\chi) 
 \bar{Y}^{\ell m,\pm} (\chi) .
\end{split}
\end{align}

Returning to our ongoing example, we can form the odd-parity, 
$(\ell,m) = (2,1)$ contribution to the $t,\vp$ and $r,\vp$ MP components.
Note that the $t,\th$ and $r,\th$ components vanish.
Using our expression from Eqn.~\eqref{eqn:psiBarOdd} in 
Eqns.~\eqref{eq:ReconstructOddBar} and \eqref{eqn:mpBarlm}
we find
\begin{align}
\begin{split}
p_{t \vp}^{21,+}
& = 
\mu  \Big[-1 - e\cos \chi+\mathcal{O}\left(e^2\right)\Big] y^{1/2}
+ 
\mu
\left[-\frac{47}{84}
+ \left(\frac{5}{12} \cos \chi+\frac{5 i }{28} \sin \chi\right)  e
+\mathcal{O}\left(e^2\right)\right]
y^{3/2}
+\mathcal{O}\left(y^{5/2}\right), \\
p_{r \vp}^{21,+}
& = \mu  \Big[-i  +  \big(\sin \chi-2 i \cos \chi\big) e
+\mathcal{O}\left(e^2\right)\Big] y \\
& \hspace{30 ex}
+ 
\mu
\left[
-\frac{271 i  }{84}
+  \left(\frac{71}{42}\sin \chi-\frac{177 i}{28} \cos \chi\right)e
+\mathcal{O}\left(e^2\right)\right]
y^{2}
+\mathcal{O}\left(y^{3}\right).
   \end{split}
\end{align}

Finally, we sum over $m$-modes. In practice we do this by taking twice the
real part of each non-zero $m$-mode and adding to the $m=0$ mode. 
As mentioned in Sec.~\ref{sec:formalism},
while RWZ gauge is known to have discontinuous MP amplitudes for 
each $\ell m$ mode, we find that these discontinuities cancel out after summing
over $m$-modes. In fact, using the expressions in Ref.~\cite{HoppEvan10}, we 
find that the Dirac delta behavior of the MP amplitudes also cancels out,
so that RWZ gauge is $C^0$ for all $\ell \ge 2$.
For the specific example of $\ell = 2$, we find the $t,\vp$
and  $r,\vp$ components
of the MP to be
\begin{align}
\label{eqn:MPl2}
\begin{split}
p_{t \vp}^2 
&=
\mu
\left[ \Big(-2-2 e \cos \chi
+\mathcal{O}\left(e^2\right)\Big) y^{1/2}
+ 
\left(-\frac{47}{42}+\frac{5}{6} e \cos \chi
+\mathcal{O}\left(e^2\right)\right) y^{3/2}
+
\mathcal{O}\left(y^{5/2}\right) 
\right],
\\
p_{r \vp}^2 
& =
\mu 
\left[  \Big(2 e \sin \chi
+\mathcal{O}\left(e^2\right)\Big) y
+
\left(\frac{71}{21} e\sin \chi
+\mathcal{O}\left(e^2\right)\right) y^2 
+\mathcal{O}\left(y^{3}\right) 
\right] .
\end{split}
\end{align}

\subsection{Large$-\ell$ expressions}

In the previous subsections, we used the example of $\ell=2$ and odd-parity
to show how we construct retarded solutions to each MP component 
for that specific $\ell$-mode. Our example only showed an expansion
through 1PN. In reality, for such a low-order expansion, we need not 
specify an explicit $\ell$ value. In fact, through 2PN we can write
down solutions for arbitrary $\ell \ge 2$. For the purposes of this work,
wherein we expand to 4PN, we calculated explicit solutions for all modes
$\ell \le 3$. For all higher modes we use generic-$\ell$ expressions.
These expressions follow from a nearly identical calculation
to the specific-$\ell$ case. In fact, it is only the 
homogeneous solutions that are different. We give an abbreviated overview
of the procedure.

In the odd parity, through 1PN the homogeneous solutions are of the form
\cite{BiniDamo13, KavaOtteWard15}
\begin{align}
\begin{split}
\label{eqn:genOddHomog}
\hat{X}_\ell^{o,+}
&=
\left(\eta ^2 \mathcal{X}_1\right)^{\ell} 
\left[1+\left(\frac{\mathcal{X}_2}{4 \ell -2}+\frac{\mathcal{X}_1 (\ell -1) (\ell +3)}{\ell +1}\right) \eta ^2
+
\mathcal{O}\left(\eta ^4\right)\right],
\\
\hat{X}_\ell^{o,-}
&=
\left(\eta ^2 \mathcal{X}_1\right)^{-\ell -1} 
\left[1
+
\left(
\frac{4 - \ell^2}{\ell} \mathcal{X}_1 
-
\frac{\mathcal{X}_2}{4 \ell +6}
\right) \eta ^2
+
\mathcal{O}\left(\eta ^4\right)\right].
\end{split}
\end{align}
Note that when $\ell=2$, these reduce
to the expressions in Eqn.~\eqref{eqn:xOdd2}.
We take these expressions, evaluate them along the particle's worldline,
then normalize them by performing the integral \eqref{eqn:CFromGFBar}.
For the even parity, the procedure is equivalent, after using 
Eqn.~\eqref{eqn:chandra} to form the 
homogeneous solutions.
Critically, during the entire calculation we keep the leading term
[either $\left(\eta ^2 \mathcal{X}_1\right)^{\ell} $ or 
$\left(\eta ^2 \mathcal{X}_1\right)^{-\ell -1} $] factored 
out of the PN expansion. This term eventually cancels out once the FD EHS are formed.
Following the procedure through, we find that the infinity-side odd-parity master 
function, is
\begin{align}
\bar{\Psi}_{\ell m}^{o,+} 
&=
\frac{16 \pi \mu \ \pa_\th Y_{2 m}(\pi/2, 0)}{(\ell -1) \ell  (\ell +1) (2 \ell +1)}
\Bigg\{
2 \Big[1 +\big( \cos \chi - 2 i m \sin \chi  \big) e 
+ \mathcal{O}\left(e^2\right) \notag
\Big] y^{1/2}\\
& \hspace{3ex} +
\frac{1}{\ell  (\ell +1) (\ell +2) (2 \ell -1) (2 \ell +3)}
\bigg[
   5 \left(2 m^2+5\right) \ell ^2+2 \left(5 m^2+13\right) \ell +12 \ell ^5+32 \ell ^4+19 \ell ^3-24 \\
&  \hspace{6ex} +
\Big(3 
\big(\left(10 m^2+19\right) \ell ^2+2 \left(5 m^2+13\right) \ell +4 \ell ^5+4 \ell ^4-7 \ell ^3-16\big) \cos \chi \notag
\\
& \hspace{6ex} -
2 i m \big(2 \left(5 m^2+18\right) \ell ^2+\left(10 m^2+17\right)
   \ell +28 \ell ^5+96 \ell ^4+87 \ell ^3-24\big) \sin \chi\Big) e
   + \mathcal{O}\left(e^2\right)
\bigg] y^{3/2}
+
 \mathcal{O}\left(y^{5/2}\right)
\Bigg\} .\notag
\end{align}
Again, this reduces to the specific expression given in Eqn.~\eqref{eqn:psiBarOdd}
when $\ell=2$.
We use the master function expressions to form the MP contributions
for each $\ell m$ mode. In order to perform the $m$ sum for generic $\ell$,
we make use of the App.~F procedure of Nakano et al.~\cite{NakaSagoSasa03}.
Our final expressions for $p_{\mu \nu}^\ell$ are again a double
expansion in $y$ and $e$. The generic-$\ell$ equivalents of
Eqn.~\eqref{eqn:MPl2} are
\begin{align}
p_{t \vp}^\ell &=
\mu \Bigg\{ 
\Big[-2-2  e \cos \chi  +  \mathcal{O}\left(e^2\right) \Big] y^{1/2} \notag
\\
& \hspace{1 ex}
+ 
\left[
-\frac{3 \left(2 \ell ^4+4 \ell ^3+7 \ell ^2+5 \ell -8\right)}{2 \ell (\ell +1) (2 \ell -1) (2 \ell +3)} 
+
\frac{14 \ell ^4+28 \ell ^3-43 \ell ^2-57 \ell +48}{2 \ell  (\ell +1) (2 \ell -1) (2 \ell +3)} 
 e \cos \chi
+ 
\mathcal{O}\left(e^2\right) \right] y^{3/2}
+
 \mathcal{O}\left(y^{5/2}\right)  \Bigg\},  \\
p_{r \vp}^\ell
&=
\mu \Bigg\{
\Big[ 2 e \sin \chi
+ \mathcal{O}\left(e^2\right)\Big] y
+
\left[
\frac{11 \ell ^4+22 \ell ^3+20 \ell ^2+9 \ell -24}{\ell  (\ell +1) (2 \ell -1) (2 \ell +3)}
e \sin \chi
+
\mathcal{O}\left(e^2\right)
\right] y^2 
+\mathcal{O}\left(y^{3}\right) 
\Bigg\} .\notag
\end{align}
As before, we find the MP to be single valued for each $\ell$.
Taking into account three different cases: low-order modes (App.~\ref{sec:lowOrder}),
specific-$\ell$ values (in our case $\ell = 2, 3$), and generic-$\ell$ for all
the rest, the full retarded MP is formed from a simple sum over $\ell$,
\begin{align}
p_{\mu \nu} = \sum_{\ell} p_{\mu \nu}^{\ell} .	
\end{align}

\subsection{Re-summation of the small eccentricity expansions}

For our work here, we used \emph{Mathematica} to 
expand all quantities in the small-$e$ limit, keeping powers up to $e^{10}$.
As eccentricity order increases, the task of simplifying large expressions
requires substantial computational resources. Indeed, finding the 
generic-$\ell$ even-parity normalization coefficient (our most taxing calculation)
took some 10 days and 20 GB of memory. Nonetheless, the virtue of this approach is 
that we need only perform that calculation once at each $\ell$.
Still, with such computational overhead, the question remains: can
this approach be useful when considering high-eccentricity orbits?
Inspired by recent work of Forseth et al. \cite{ForsEvanHopp15}
(see also Ref.~\cite{Evan15a}), we have sought to ``re-sum''
our final results at each PN order so as to capture the $e\to 1$ behavior. 
(We note also that it is likely possible, and perhaps simpler, to achieve
the same result by expanding in $p^{-1}$ instead of $y$ \cite{Bara15,Evan15b}.)
As an example, consider the 0PN expression for $p_{t \vp}^2$, shown here
with all 10 powers of eccentricity,
\begin{align}
\begin{split}
p_{t \vp}^2 & =
- \mu \Bigg[
2 
+
2 e \cos \chi
+
e^2
+
e^3  \cos \chi
+
\frac{3}{4} e^4
+
\frac{3}{4} e^5 \cos \chi
+
\frac{5}{8} e^6 \\
& \hspace{30 ex}
+
\frac{5}{8} e^7 \cos \chi 
+
\frac{35}{64} e^8 
+
\frac{35}{64} e^9 \cos \chi
+
\frac{63}{128} e^{10}
+
\mathcal{O}\left(e^{11}\right)
\Bigg] y^{1/2}
+
\mathcal{O}\left(y^{3/2}\right) .
\end{split}
\end{align}
It is a relatively simple task to guess that this series is probably
the small-$e$ expansion of
\begin{align}
p_{t\vp}^{2} &= -\frac{2 \mu  (1 + e \cos \chi)}{(1-e^2)^{1/2}} y^{1/2}
+
\mathcal{O}\left(y^{3/2}\right) .
\end{align}
A similar analysis of our generic-$\ell$ solutions provided closed-form
expressions for all the 0PN and 1PN retarded MP components.

Beyond 1PN, finding closed-form solutions becomes harder.
Examining the PN literature on eccentric orbits 
(see, e.g. Eqn.~(356) of Blanchet's Living Review \cite{Blan13}), 
it is clear that starting
at 2PN the $e \to 1$ singular behavior becomes more subtle than some simple
inverse factor of $1-e^2$. And yet, as a first approximation,
factoring out the appropriate leading power of $1-e^2$ at each PN order
[each order comes with an additional factor of $(1-e^2)^{-1}$],
yields dramatically improved convergence for high eccentricities.

We note, of course, that our inability to find closed-form expressions
beyond 1PN does not imply their non-existence. Indeed, from standard PN
calculations, the metric of two bodies in eccentric motion is
known through 3PN with arbitrary mass-ratios. See, Ref.~\cite{AkcaETC15}
and references therein where the metric is given in standard harmonic
coordinates. Here we simply present the results we were able to deduce 
by re-summing our small-eccentricity expansion, and without making use
of known results from other research.

\section{The generalized redshift invariant}

\label{sec:UAvgBackground}

Before continuing to our results, we introduce the specific gauge invariant
that we will compute. We first briefly cover some GSF background and then
give the exact expression we use in our calculations.

\subsection{Abbreviated background on gravitational self-force invariants}

When working at zeroth order in the mass-ratio  $q$, 
the particle moves on a geodesic of the background spacetime $g_{\mu \nu}$, 
as we have assumed up to this point. Once we allow the particle to 
have a small, but finite mass $\mu$, the motion is no longer geodesic in $g_{\mu \nu}$,
but it is geodesic in the effective spacetime 
${\rm g}_{\mu \nu} = g_{\mu \nu} + p^R_{\mu \nu}.$
Here $p^R_{\mu \nu}$ is the \emph{regular} part of the MP, due to 
Detweiler and Whiting \cite{DetwWhit03}. The regular MP comes from removing the particle's own
singular field $p^S_{\mu \nu}$ from the retarded MP $p_{\mu \nu}$. 
An appropriate gradient of $p^R_{\mu \nu}$ provides the GSF; see e.g. \cite{BaraSago10,OsbuETC14}.

When computing local gauge invariants, 
it is convenient to ``turn off'' the dissipation
due to the GSF and look at effects due solely to the conservative GSF. This is 
done by forming 
$p_{\mu \nu}^{R,\rm{cons}} =  \l p_{\mu \nu}^{R,\rm{ret}} + p_{\mu \nu}^{R,\rm{adv}} \r / 2$.
Here $p_{\mu \nu}^{R,\rm{ret}}$ is the regular MP computed with retarded boundary 
conditions, while $p_{\mu \nu}^{R,\rm{adv}}$ is computed with advanced boundary 
conditions. All gauge-invariant GSF quantities mentioned in the introduction
are defined with respect to the conservative effective spacetime
${\rm g}^{\rm cons}_{\mu \nu} = g_{\mu \nu} + p^{R,{\rm cons}}_{\mu \nu}.$
However, when computing an orbit average, the dissipative contribution 
to ${\rm g}^{\rm cons}_{\mu \nu}$ averages to zero, and so we can compute 
such quantities from directly ${\rm g}_{\mu \nu}$.

\subsection{Practical details of the $\langle U \rangle$ calculation}

We are now ready to calculate the  so-called ``generalized redshift invariant''. 
This quantity is an eccentric orbit extension
of Detweiler's \cite{Detw08}
original invariant $u^t$. The generalization $\langle U \rangle$
is due to Barack and Sago \cite{BaraSago11}, and is the orbit average of $u^t$ with
respect to proper time $\tau$. 
Critically, this average is taken on the GSF-\emph{perturbed} orbit. 
In the language introduced above, the average is performed along the 
geodesic motion in the effective spacetime ${\rm g}_{\mu \nu}$.
When mapping between the background
metric $g_{\mu \nu}$ and the effective metric ${\rm g}_{\mu \nu}$,
we take $\O_i$ to be fixed. This implies that $T_r$ has the same
value in the two spacetimes, but the $\tau$ radial period $\mathcal{T}_r$ does not.
Thus, we write the orbit average of $u^t$ with respect to proper time as
\begin{align}
\langle U \rangle \equiv 
\langle u^t \rangle 
= \frac{T_r}{\mathcal{T}_r + \d \mathcal{T}_r},
\end{align}
where $\d \mathcal{T}_r$ is a $\mathcal{O} (q)$ shift 
in $\mathcal{T}_r$ due to the conservative GSF, 
(Note that here we use $T_r$ and $\mathcal{T}_r$ to indicate background quantities,
a notation which differs from \cite{AkcaETC15}.)

Before continuing we wish to emphasize a point stressed in 
Refs.~\cite{BaraSago11,AkcaETC15}.
It is not enough to simply compute a gauge-invariant quantity like
$\langle U \rangle$. Rather, one must find a functional way to relate that quantity
to the particular orbit being considered. A natural gauge-invariant characterization 
of perturbed orbits is the two fundamental frequencies, $\O_r$ 
and $\O_\vp$ 
(recall that $\d \O_r = \d \O_\vp = 0$ due to frequency fixing).
If one can parametrize a GSF-perturbed orbit with observable frequencies,
then the nontrivial gauge invariant $\langle U \rangle$ can be taken as a function
of those two parameters. In this way, it is possible for those working in 
different gauges (such as those used in PN literature) to compare results 
in a meaningful way. 

We now expand $\langle U \rangle$ into a background part, and
a part due to the GSF,
\begin{align}
\langle U \rangle \l \O_i \r 
= \langle U \rangle_0 \l \O_i \r
+ q \langle U \rangle_{\rm gsf} \l \O_i \r,
\q \q \langle U \rangle_0 = \frac{T_r}{\mathcal{T}_r},
\end{align}
where $\O_i \equiv \left\{ \O_r, \O_\vp \right\} $.
When $\O_i$ are taken to be fixed,
Akcay et al.~\cite{AkcaETC15} show that the $\mathcal{O} (q)$ shift  in
$\langle U \rangle$ can be reduced to
\begin{align}
\label{eqn:UAvgFinal}
q \langle U \rangle_{\rm gsf}
= 
- \frac{\langle U \rangle_0}{\mathcal{T}_r} \, \d \mathcal{T}_r
 =
\frac{T_r}{\mathcal{T}_r^2}
\left\langle H^R \right\rangle,
\end{align}
a simplification over the original Barack and Sago expression.
Here we follow the notation of 
Refs.~\cite{BaraSago11,HeffOtteWard12a} and introduce
\begin{align}
\label{eqn:HR}
H^R \equiv \frac{1}{2} p^R_{\mu \nu} u^\mu u^\nu 
= \frac{1}{2} p_{\mu \nu} u^\mu u^\nu -H^S,
\end{align}
where $H^S$ is due to the singular field $p^S_{\mu \nu}$.

The calculation of the singular field is a subtle task.
Its removal is most often performed with mode-sum regularization 
\cite{BaraOri00}, wherein singular terms are subtracted $\ell$-by-$\ell$
(though other techniques exist; see Ref.~\cite{Ward15}
for an overview). Barack and Sago first computed 
the leading-order regularization parameter for $H^S$ when they
introduced $\langle U \rangle$.
Since then, higher-order regularization parameters 
(improving convergence rates for cases when the $\ell$ sum must be truncated)
have been computed
by Heffernan et al.~\cite{HeffOtteWard12a}. But, for our 
purposes, where we know all $\ell$, 
we only need the leading-order term,
\begin{align}
\label{eqn:HS}
H^S = \sum_\ell H_{[0]} \equiv \sum_\ell \frac{2 \mu}{\pi \sqrt{\mathcal{L}^2+r_p^2}} 
\mathcal{K} \l \frac{\mathcal{L}^2}{\mathcal{L}^2+r_p^2} \r.
\end{align}
The 
$H_{[0]}$ regularization parameter notation follows Ref.~\cite{HeffOtteWard12a};
the subscript $[0]$ indicates that $H_{[0]}$ multiplies zero powers of $\ell$.
Combining Eqn.~\eqref{eqn:HS} with \eqref{eqn:HR} gives
\begin{align}
\left\langle H^R \right\rangle= 
\sum_\ell \l \frac{1}{2} \left\langle p^\ell_{\mu \nu} u^\mu u^\nu \right\rangle
- \left\langle H_{[0]} \right\rangle \r.
\end{align}
For our present calculation, it is a simple task to use our expansions for 
$r_p$ and $\mathcal{L}$ and obtain a PN expansion for $H_{[0]}$.
Both $p^\ell_{\mu \nu} u^\mu u^\nu $ and $H_{[0]}$ are then
averaged over one $\tau$-period with the use of the PN expansion for
$d\tau_p / d \chi$ from Eqn.~\eqref{eqn:darwinEqns}, thus forming 
$\left\langle p^\ell_{\mu \nu} u^\mu u^\nu \right\rangle$ 
and $\left\langle H_{[0]} \right\rangle$,
and forming all that we need for a practical calculation of 
$\langle U \rangle_{\rm gsf}$.

Lastly, we note that Akcay et al.~\cite{AkcaETC15} adjust for the 
non-asymptotic-flatness of the Lorenz gauge monopole by adding a correction 
term to Eqn.~\eqref{eqn:UAvgFinal}. In App.~\ref{sec:lowOrder}
we show that the original RWZ monopole \cite{Zeri70} is also not asymptotically flat,
but we are able to correct that with a slight gauge transformation,
and so we use Eqn.~\eqref{eqn:UAvgFinal} exactly as is.
However, we note that the radiative modes of RWZ gauge are \emph{not} asymptotically
flat \cite{GleiETC00,PresPois06,HoppEvan13}. This is curious, for, Barack and Sago
established the gauge invariance of $\langle U \rangle_{\rm gsf}$ for 
a certain class of gauges which respect the periodicity of the orbit
and are well behaved at spatial infinity. Still, we have empirical evidence
from several calculations (including this one) that RWZ gauge
falls into the class of gauges for which $\langle U \rangle_{\rm gsf}$ 
is invariant. The question remains, why must we correct the 
non-asymptotic-flatness of the monopole, but not other modes?
At this point the answer is not clear.

\section{Results}
\label{sec:results}

Following the procedure described above, 
we have used \emph{Mathematica} to compute the MP along with 
its first $t$ and $r$ derivatives through 4PN while keeping
powers in eccentricity up to $e^{10}$. Our PN order required us to compute 
solutions to the modes $\ell=2$ and $\ell=3$ explicitly, while all
higher $\ell$ modes are described by a general-$\ell$ expression.

For each specific-$\ell$, as well as the generic-$\ell$ case, 
we performed the calculation for both even and odd parities,
on both sides of the particle. After summing over $m$-modes
we noticed the surprising result that RWZ gauge is in fact $C^0$ 
for each $\ell$, despite being to be discontinuous with delta functions
at the $\ell m$ level. We subsequently confirmed that this was true to all
PN orders using the expressions in Ref.~\cite{HoppEvan10}.

\begin{center}
\begin{figure}[t!]
\includegraphics[scale=1]{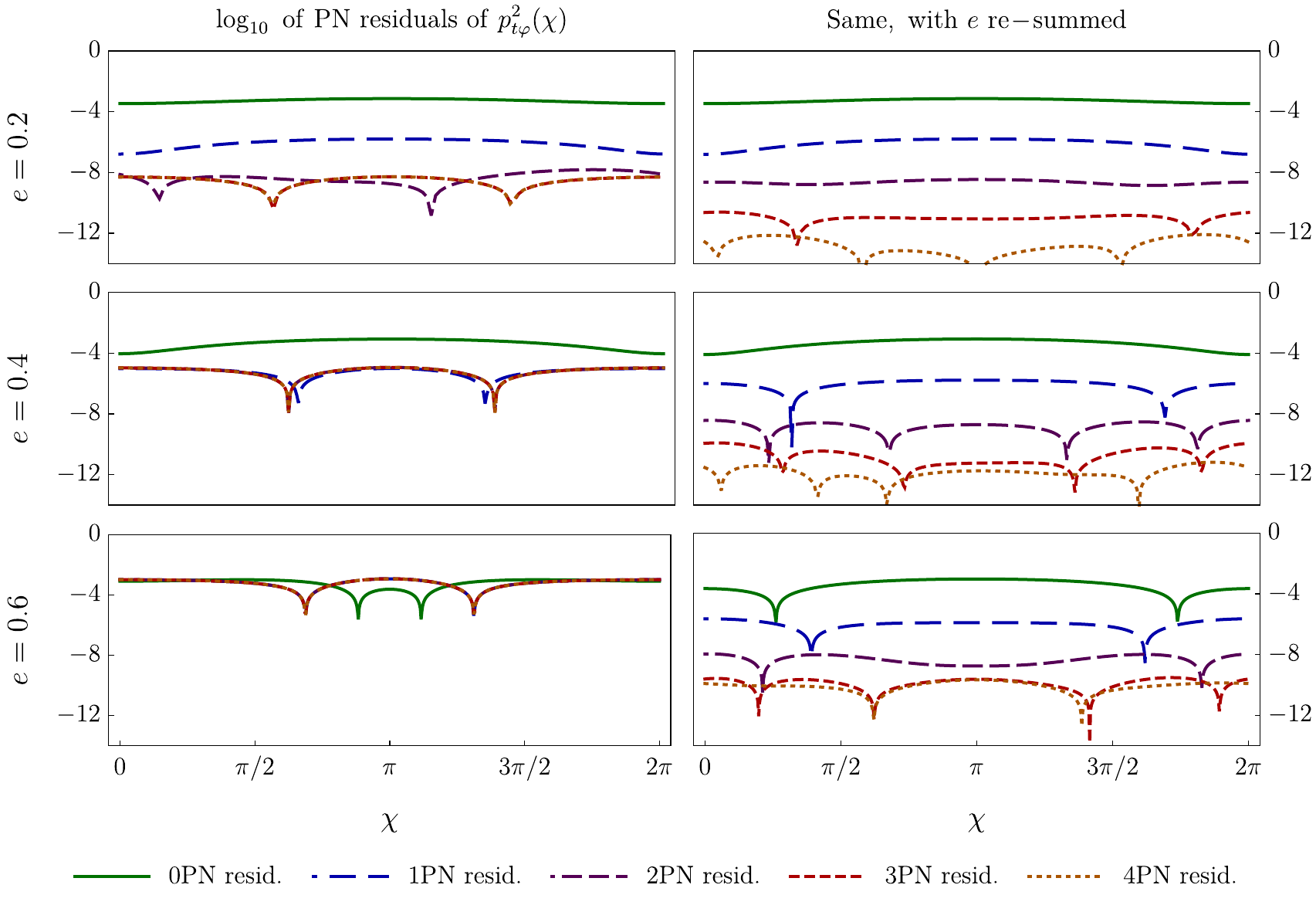}
\caption{
The effect of re-summing the small-$e$ expansion at each PN order
as seen by comparing to numerical data, all computed at $p=1000$.
We see that especially as eccentricity increases, our re-summation
greatly improves convergence. Also, note the consistency of our convergence
throughout the orbit. Our results are no less effective at periapsis
than apoapsis. 
The dips in the residuals are from zero crossings, and not meaningful.
See the discussion in the text for more details.
\label{fig:panelPlot}} 
\end{figure}
\end{center}

We show the convergence of one of the MP components with PN order
in Fig.~\ref{fig:panelPlot}.
We used a numerical code developed for recent work \cite{ForsEvanHopp15}
to compute the $\ell=2$ contribution to the MP component $t$,$\vp$
at three different eccentricities.
We then subtracted successive PN terms derived analytically for this
work and computed the relative error. 
In the left column we see that even for a moderately low eccentricity
of $e=0.2$ the PN convergence stalls at 2PN due to the small-$e$ expansion.
At $e=0.6$ the convergence stalls after the subtraction of only the 0PN term.
In the right column we see that factoring out the $e\to 1$ singular
behavior greatly improves the convergence. Note that at $e=0.6$ 
the convergence still appears to stall around 3PN. In order to probe such
eccentricities at the 4PN level, 
we will evidently need more than 10 powers of $e$, or 
perhaps a more precise capturing of the $e \to 1$ singular behavior 
for 2PN and beyond.

We now provide our results for the invariant $\langle U \rangle_{\rm gsf}$,
computed using the procedure described in Sec.~\ref{sec:UAvgBackground}.
The specific expression is given below in Eqn.~\eqref{eqn:UAvgY}
($\gamma$ is the Euler-Mascheroni  constant).
We performed the regularization in two ways so as to check
our removal of the singular field. First, we fit out the constant-with-$\ell$
term by taking the large-$\ell$ limit of our generic-$\ell$ expressions.
That fit-out regularization parameter exactly agreed with the proper time
average of $H_{[0]}$ when expanded in $y$ and $e$. Using \emph{Mathematica}
we are able to take the $\ell$ sum all the way to infinity, and thus have
no error due to truncation.

\begin{center}
\begin{figure}[t!]
\includegraphics[scale=1]{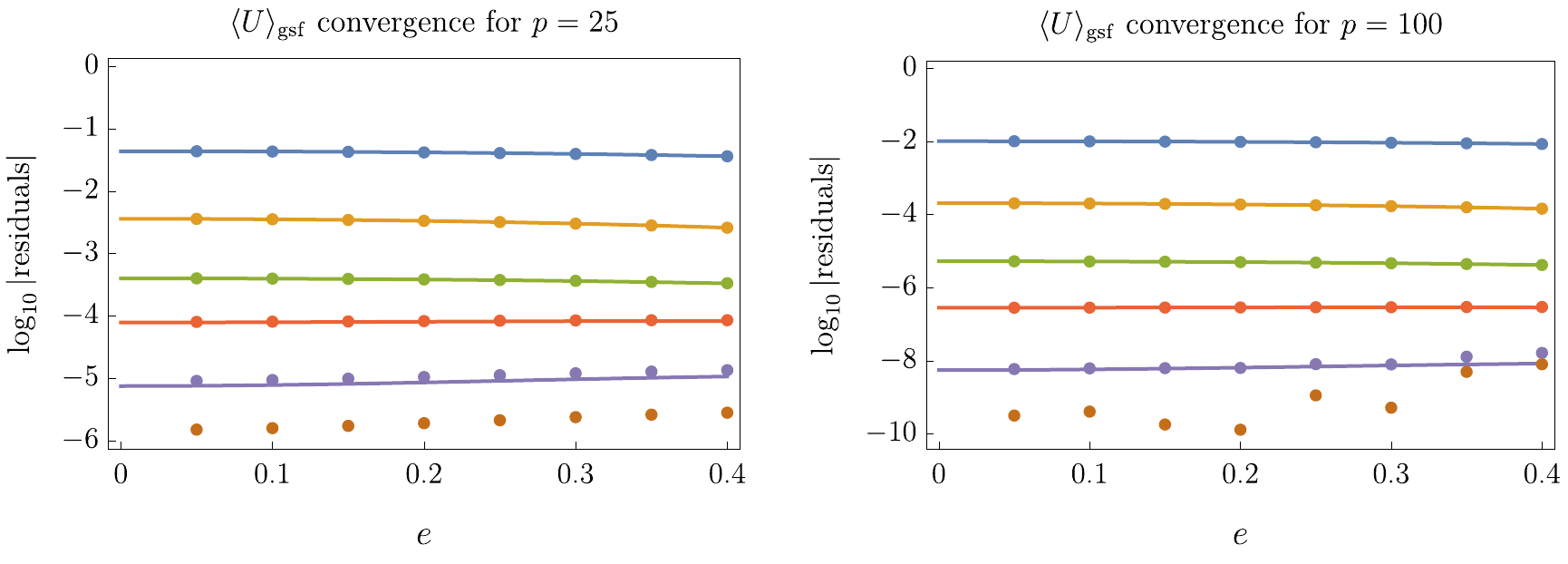}
\caption{
Comparison of our PN expressions with numerical data from 
Table II of Akcay et al.~\cite{AkcaETC15}.
Dots are numerical values and lines are our analytic calculations. The top
row is $\log_{10}$ of the absolute value of the full $\langle U \rangle_{\rm gsf}$.
Successive rows show residuals after subtracting each term in the 
PN series. 
The consistency of our agreement out to $e=0.4$ is only possible because of our
re-summation of the small-$e$ expansion.
Unevenness of the final residuals for $p = 100$ is likely
a numerical artifact.
\label{fig:UConv}} 
\end{figure}
\end{center}

We have compared our expression to the published 3PN values of 
Akcay et al.~\cite{AkcaETC15},
which were computed by starting in standard harmonic coordinates.
That reference also provides numerical data for $\langle U \rangle_{\rm gsf}$,
computed in Lorenz gauge, which we compare to in Fig.~\ref{fig:UConv}.
The recent RWZ gauge work by Bini et al.~\cite{BiniDamoGera15},
provides an analytic 4PN, $\mathcal{O} \l e^2 \r$ value of 
$\langle U \rangle_{\rm gsf}$, which we agree with as well.
Lastly, van de Meent and Shah \cite{VandShah15}, who work in radiation
gauge, provide numerical predictions for 4PN terms through $e^6$ and we agree
with all of their values within the provided error bars.

\begin{align}
\label{eqn:UAvgY}
\begin{split}
\langle U \rangle_{\rm gsf} &=
-y
-
\frac{2 \left(1-2 e^2\right)}{1-e^2} y^2
+
\frac{1}{\left(1-e^2\right)^2}
\left[
-5+16 e^2-\frac{85 e^4}{8}-\frac{9 e^6}{16}-\frac{17 e^8}{128}-\frac{7 e^{10}}{256}
 +\mathcal{O}\left(e^{12}\right) 
\right] y^3
 \\
& \hspace{3 ex} +
\frac{1}{\left(1-e^2\right)^3}
\Bigg[
-\frac{121}{3}+\frac{41 \pi ^2}{32}+36 e^2
+e^4 \left(-\frac{453}{8}-\frac{123 \pi ^2}{256}\right)
+e^6 \left(\frac{251}{8}-\frac{41 \pi ^2}{256}\right)\\
& \hspace{32 ex} 
+e^8 \left(\frac{909}{128}-\frac{369 \pi^2}{4096}\right)
+e^{10} \left(\frac{217}{64}-\frac{123 \pi ^2}{2048}\right)
+\mathcal{O}\left(e^{12}\right) 
\Bigg] y^4
 \\
& \hspace{3 ex} +
\frac{1}{\left(1-e^2\right)^4} 
\left[ -\frac{64}{5}-\frac{488 e^2}{15}
+\frac{242 e^4}{15}+\frac{122 e^6}{15}
+\frac{71 e^8}{20}+\frac{523 e^{10}}{240}
+\mathcal{O}\left(e^{12}\right) \right] y^5 \log y
 \\
& \hspace{3 ex} +
\frac{1}{\left(1-e^2\right)^4}
\Bigg[
-\frac{1157}{15}-\frac{128 \gamma}{5}+\frac{677 \pi ^2}{512}-\frac{256}{5} \log 2 \\
& \hspace{15 ex} 
+
e^2 \left(\frac{13102}{45}-\frac{976 \gamma}{15}+\frac{2239 \pi ^2}{3072}
+\frac{496}{3} \log 2-\frac{1458}{5} \log 3\right)\\
& \hspace{15 ex} 
+
e^4 \left(-\frac{7679}{90}+\frac{484 \gamma}{15}-\frac{21941 \pi ^2}{6144}
-\frac{20724}{5} \log 2+\frac{5103}{2} \log 3\right)\\
& \hspace{15 ex} 
+
e^6 \left(\frac{21859}{120}+\frac{244  \gamma}{15}+\frac{13505 \pi ^2}{12288}
+\frac{157564}{5} \log 2-\frac{1586061}{160} \log 3-\frac{1953125}{288} \log 5\right)\\
& \hspace{15 ex} 
+
e^8 \left(-\frac{1937767}{23040}
+\frac{71 \gamma}{10}
+\frac{13129 \pi  ^2}{65536}
-\frac{13131809}{90} \log 2
-\frac{1355697}{1280} \log 3
+\frac{48828125}{768} \log 5\right)\\
& \hspace{15 ex} 
+
e^{10} \bigg(-\frac{3734707}{115200}
+\frac{523 \gamma}{120}
+\frac{31055 \pi^2}{393216}
+\frac{14107079851}{27000} \log 2
+\frac{202776271479}{1024000} \log 3\\
& \hspace{35 ex} 
-\frac{59986328125}{221184} \log 5-\frac{678223072849}{9216000} \log 7\bigg)
+\mathcal{O}\left(e^{12}\right) 
\Bigg] y^5 +\mathcal{O}\left(y^6\right) .
\end{split}
\end{align}

We note that (with an eye to the discussion in
Sec.~\ref{sec:UAvgBackground}), 
our parametrization \eqref{eqn:UAvgY}
is somewhat lacking. Since $y$ is derived from $\O_\vp$, it
is gauge-invariant, but $e$, however, is not. When
considering eccentric orbits, it is customary to give results in 
terms of $y$ and $\la \equiv  3y / k$, where $k \equiv \O_\vp / \O_r - 1$
is a measure of periapsis advance. 
It is a straightforward, if tedious, task
to transform our results to $\lambda$, but in the process we lose the 
physical intuition that $e$ provides. 
In practice it is more convenient to convert from $\la$ to $e$ for the 
purposes of comparison.
In App.~\ref{sec:UAvgP},
we do provide $\langle U \rangle_{\rm gsf}$ as a function of $p$ and $e$,
though for easy comparison with GSF codes.

In addition to comparing with other calculations of $\langle U \rangle_{\rm gsf}$,
our result is useful for EOB comparison and calibration. Notably,
recent 4PN EOB results
have introduced terms through $e^6$ in the non-geodesic ``$\hat Q$ potential''.
See Eqn.~(8.1c) of Damour et al.~\cite{DamoJaraScha15}. 
The full $\hat Q$ can be separated into terms linear-in-$\nu$ (with $\nu$ being
the symmetric mass-ratio), which
are accessible through our GSF calculation, and terms which are 
$\mathcal{O} \l \nu^2 \r$ and beyond, which would require at least 
the second-order GSF to compute. Thus, we write
$\hat Q \l u , p_r'\r = \nu q \l u , p_r'\r + \mathcal{O} \l \nu^2 \r$.
(The details of the EOB notation used here and throughout the rest of this section
can be found in Ref.~\cite{DamoJaraScha15}.)
Our expression for $q \l u , p_r'\r$, through $e^{10}$ [equivalently
$(\mathbf{n}' \cdot \mathbf{p}')^{10}$] is
%
\begin{align}
q(u,p_r') &= 
8 u^2 (\mathbf{n}' \cdot \mathbf{p}')^4+ 
\l -\frac{5308}{15} + \frac{496256}{45} \log 2 - \frac{33048}{5} \log 3 \r 
u^3 (\mathbf{n}' \cdot \mathbf{p}')^4 \notag \\
& \hspace{-2ex} + 
\l -\frac{827}{3} - \frac{2358912}{25} \log 2 + \frac{1399437}{50} \log 3  
+ \frac{390625}{18} \log 5\r 
u^2 (\mathbf{n}' \cdot \mathbf{p}')^6 
\label{eqn:qPotential} \\
& \hspace{-2ex} + 
\l  -\frac{35772}{175}+\frac{21668992}{45}  \log 2
 +\frac{6591861}{350} \log 3
 -\frac{27734375}{126} \log 5 \r
u (\mathbf{n}' \cdot \mathbf{p}')^8 \notag \\
& \hspace{-2ex} + 
\l
-\frac{231782}{1575}
-\frac{408889317632}{212625}  \log 2
-\frac{22187736351}{28000} \log 3 
+\frac{7835546875}{7776} \log 5
+\frac{96889010407}{324000} \log 7
\r (\mathbf{n}' \cdot \mathbf{p}')^{10} \notag \\
& \hspace{-2ex} + 
\mathcal{O} \left[ u^{-1} (\mathbf{n}' \cdot \mathbf{p}')^{12}\right] .\notag
\end{align}
We derived Eqn.~\eqref{eqn:qPotential} using the procedure described in
Le Tiec's recent work \cite{Leti15}. Using Eqn.(5.27) from that reference,
we were able to transcribe our 4PN, $e^4$ contribution to $\langle U \rangle_{\rm gsf}$
to the  $u^3 (\mathbf{n}' \cdot \mathbf{p}')^4$ contribution to $q$,
confirming the same term in Eqn.~(8.1c) of Ref.~\cite{DamoJaraScha15}.
We then followed the prescription of Le Tiec to derive 
transcription equations equivalent to his (5.27) for $e^6$, $e^8$, and $e^{10}$.
With these, we confirmed the $u^2 (\mathbf{n}' \cdot \mathbf{p}')^6$ coefficient 
(the $e^6$ term) of $q$, given in Ref.~\cite{DamoJaraScha15}. 
The last two terms in Eqn.~\eqref{eqn:qPotential} are previously 
unknown coefficients, corresponding to $e^8$ and $e^{10}$ at 4PN.

\section{Conclusions and Outlook}
\label{sec:outlook}

We have presented a method for solving the first-order field 
equations in a PN/small-eccentricity 
expansion when the source is a point particle in bound
motion on a Schwarzschild background. In this work we have kept
terms through 4PN and $e^{10}$, but our method will extend 
naturally to higher orders. Important to the effectiveness of our results
was the re-summing of the $e$-series at each PN order.
Our method lends itself to many further calculations of eccentric
orbit invariants. Since we already have computed
derivatives of the MP (though they were not used in computing 
$\langle U \rangle$ here), 
a natural next step is to compute an eccentric
orbit generalization of the spin-invariant $\psi$ \cite{DolaETC14a}.

Moving beyond Schwarzschild to Kerr is a more challenging task. 
The analytic merger of PN theory with black hole perturbation theory on Kerr
has a long history (e.g., \cite{GanzETC07,SagoFuji15}), though
the focus has typically been on fluxes and nonlocal dissipative GSF. 
The reason is largely due to the challenge of reconstructing the
radiation-gauge MP from the Teukolsky variable. 
Recent work by Pound et al.~\cite{PounMerlBara13}
has helped to clarify the subtleties of the process, and it may now
be possible to extend our method to compute the Kerr MP, although
the task is formidable.

\acknowledgments

We thank Barry Wardell for suggestions and assistance with regularization.
Charles Evans, and Leor Barack were helpful in providing suggestions regarding
the $e$ re-summation, for which we are grateful. 
We thank Thibault Damour and Alessandro Nagar for 
bringing the comparison of the EOB $q(u)$ potential to our attention.
And, we are thankful for helpful comments on an earlier version of this paper
provided by Sarp Akcay.
We acknowledge support from Science Foundation Ireland under Grant 
No.~10/RFP/PHY2847. SH is grateful for hospitality 
of the Institut des Hautes \'Etudes Scientifiques, where part of this research
was conducted.
SH acknowledges financial support provided under the European 
Union’s H2020 ERC Consolidator Grant “Matter and strong-field gravity: 
New frontiers in Einstein’s theory” grant agreement no. MaGRaTh–646597.
CK is funded under the Programme for Research in Third Level 
Institutions (PRTLI) Cycle 5 and co-funded under 
the European Regional Development Fund.

\appendix

\section{Low-order modes}

In this appendix we give the Zerilli's \cite{Zeri70} analytic
solutions to the $\ell=0,1$ equations. We find that we must shift
the monopole, as is done in Lorenz gauge \cite{SagoBaraDetw08} in
order to find an asymptotically flat solution.
It is straightforward to take the expressions given here and expand
them at the particle's location in a PN series using 
the expressions in Sec.~\ref{sec:pnOrb}.

\label{sec:lowOrder}
\subsection{Monopole}

In the monopole case $\ell=m=0$ and only $h^{00}_{tt}$, 
$h^{00}_{tr}$, $h^{00}_{rr}$, and $K^{00}$ are 
defined. 
Zerilli chooses to
set $h^{00}_{tr} = K^{00} = 0$. The remaining non-zero solutions are
\begin{align}
h^{00}_{tt} = 
4 \sqrt{\pi} \mu \left[ \frac{{\cal E} }{r} 
- 
 \frac{f}{\mathcal{E} f_p r_p} \l 2 \mathcal{E}^{2} - U_{p}^2  \r 
 \right] \theta [r- r_p(t)], \q \q
h^{00}_{rr}
 =
\frac{4 \sqrt{\pi} \mu {\cal E}}{f^2 r} \theta [r- r_p(t)].
\end{align}
Recall the distinction between quantities with a subscript $p$, such as
$f_p$ which is a function of the particle location $r_p$,
and those without subscripts, 
like $f$ which is a function of the Schwarzschild coordinate $r$.
We seek asymptotically flat solutions which fall off at least as $1/r$. The amplitude
$h^{00}_{rr}$ satisfies this, but as in Lorenz gauge $h^{00}_{tt}$ does not. We therefore look
for a gauge transformation to remove the term 
\begin{align} 
- 4 \sqrt{\pi} \mu  f \frac{2 \mathcal{E}^{2} - U_{p}^2}{\mathcal{E} f_p r_p}  \theta [r- r_p(t)]	 .
\end{align}
The push equations for the three non-zero amplitudes are
\cite{MartPois05}
\begin{align}
\begin{split}
\label{eqn:pushEven}
\D h^{00}_{tt} = - 2 \pa_{t} \xi^{00}_{t} + f \frac{2M}{r^{2}} \xi^{00}_{r},\q \q 
\D h^{00}_{tr} = - f \pa_r \l f^{-1} \xi^{00}_t \r - \pa_t \xi^{00}_r, \q \q 
\D h^{00}_{rr} = - 2 \pa_{r} \xi^{00}_{r} - \frac{2M}{f r^{2}} \xi^{00}_{r}, 
\end{split}
\end{align}
From this we see
that we can take $\xi^{00}_r = 0$ and demand that
$\xi^{00}_t$ satisfy 
\begin{align} 
\Delta h^{00}_{tt} =
- 2 \pa_t \xi^{00}_t &=
4 \sqrt{\pi} \mu f
\frac{2 \mathcal{E}^{2} - U_{p}^2}{\mathcal{E} f_p r_p} .
\end{align}
Note that the Zerilli solution is zero on the horizon side of the particle,
but this push will add a nonzero term there.
As far as pushing $h^{00}_{tt}$ goes, we can 
get away with simply specifying the time derivative of 
$\xi^{00}_t$. However, a gauge transformation
involving $\xi^{00}_t$ will also affect $h^{00}_{tr}$. 
Looking at the  $h^{00}_{tr}$ push, we will need
$- f \pa_r \l f^{-1} \xi^{00}_t \r $, i.e.
\begin{align} 
\Delta h^{00}_{tr} =
- f \pa_r \l f^{-1} \xi^{00}_t \r  
&=
\frac{2 \sqrt{\pi} \mu}{\mathcal{E}} f
\pa_r \left[
\int_{0}^t
\frac{2 \mathcal{E}^{2} - U_{p}^2}{f_p r_p}  dt' \right] = 0.
\end{align}
So this choice of gauge transformation actually leaves $h^{00}_{tr}$ unchanged.
Then, multiplying by $Y^{00}$, 
in our asymptotically flat gauge the nonzero monopole solutions are
\begin{align}
p^{00}_{tt}  \l x^\mu \r &= 
2 \mu \frac{{\cal E} }{r} 
 \theta [r- r_p(t)]
+
2 \mu f
\frac{2 \mathcal{E}^{2} - U_{p}^2}{\mathcal{E} f_p r_p} 
\theta [r_p(t)- r],
 \q \q
p^{00}_{rr} \l x^\mu \r
 =
\frac{2 \mu {\cal E}}{f^2 r} \theta [r- r_p(t)].
\end{align}

\subsection{Odd-parity dipole}

When $(\ell,m)=(1,0)$ the odd-parity amplitude $h^{10}_2$ is not defined 
and we have residual gauge freedom. Zerilli
used this freedom to set $h^{10}_r=0$. The only nonzero remaining amplitude is
\begin{align}
h^{10}_t &= 
4 \mu \mathcal{L} \sqrt{\frac{\pi }{3}}  \l
 \frac{1}{r} \theta [r-r_p(t)]
+
 \frac{r^2}{r_p^3}  \theta [r_p(t)-r] \r
.
\end{align}
This amplitude decays as $r^{-1}$ at large radii and as such is asymptotically flat.
Multiplying by $X_\vp^{10}$ gives the only nonzero MP component,
\begin{align}
p_{t \vp}^{10} (x^\mu) &=
-2 \mu \mathcal{L} \sin^2 \th 
\left(
 \frac{1}{r} \theta [r-r_p(t)]
+
 \frac{r^2}{r_p^3} \theta [r_p(t)-r] 	
 \right).
\end{align}
We note that (as brought to our attention by Leor Barack),
this dipole solution is in fact singular at the horizon
(when viewed horizon-regular, ingoing Eddington-Finkelstein 
coordinates).
While this does not 
affect our first-order-in-$q$ calculation, any second-order GSF
analysis will require a horizon-regular solution.

\subsection{Even-parity dipole}

When $(\ell,m)=(1,1)$ the even-parity amplitude $G^{11}$ is not defined. 
Zerilli sets $K^{11}=0$ 
(in addition to the usual $j^{11}_t=j^{11}_r=0$).
The remaining nonzero even-parity dipole amplitudes are
\begin{align}
\begin{split}
h^{11}_{tt} &=
- 2 \mu \sqrt{\frac{2 \pi }{3}} \frac{ r f_p}{f}
\left[
\frac{\mathcal{E} r_p}{r^3}
+
\frac{6 \mathcal{L}^2 M+6 M r_p^2-3 \mathcal{L}^2 r_p+\left(2 \mathcal{E}^2-3\right) r_p^3}{\mathcal{E} r_p^5}
   -\frac{6 i \mathcal{L} \dot{r}_p}{r_p^3}
\right]
e^{-i \varphi_p (t)} 
\theta[r-r_p(t)], \\
h^{11}_{tr} &=
-\frac{2 \sqrt{6 \pi } \mu}{r f^2 r_p} 
\Big(
i \mathcal{L} f_p^2 -\mathcal{E} r_p \dot{r}_p
\Big)
 e^{-i \varphi_p (t)}
\theta[r-r_p(t)], \\
h^{11}_{rr} &=
-\frac{2 \sqrt{6 \pi } \mu  \mathcal{E} r_p f_p}{r^2 f^3} e^{-i \varphi_p (t)}
\theta[r-r_p(t)] .
\end{split}
\end{align}
We observe that $h_{tt}^{11}$ is not asymptotically flat, while
the other two amplitudes are. However, the even-parity dipole is the so-called
``pure gauge'' mode. It is a straightforward calculation to form its contribution
to $\langle U \rangle$, and we find that it vanishes in a pointwise sense.

\newpage
\section{The generalized redshift invariant as an expansion in $p^{-1}$}
\label{sec:UAvgP}

\begin{align}
\label{eqn:UAvgP}
\begin{split}
\langle U \rangle_{\rm gsf} &=
- \Big( 1-e^2 \Big) p^{-1}
 +
\Big( -2+4 e^2-2  e^4 \Big) p^{-2}
 +
\left[ -5+7 e^2+\frac{e^4}{4}-\frac{5 e^6}{2}
+\frac{15 e^8}{64}+\frac{3 e^{10}}{64} 
+
\mathcal{O}\left(e^{12}\right) \right] p^{-3} \\
& \hspace{3 ex}  +
\Bigg[ -\frac{121}{3}+\frac{41 \pi ^2}{32}
+
e^2 \left(-\frac{5}{3}-\frac{41 \pi ^2}{32}\right)
+
e^4 \left(\frac{705}{8}-\frac{123  \pi ^2}{256}\right)
+
e^6  \left(-\frac{475}{12}+\frac{41 \pi ^2}{128}\right) \\
& \hspace{33 ex}  
+
e^8 \left(-\frac{1171}{384}+\frac{287 \pi ^2}{4096}\right)
+
e^{10} \left(-\frac{115}{128}+\frac{123 \pi ^2}{4096}\right)
+
\mathcal{O}\left(e^{12}\right) 
\Bigg] p^{-4}
\\
& \hspace{3 ex}  +
\left[\frac{64}{5}+\frac{296 e^2}{15}-\frac{146  e^4}{3}+8 e^6+\frac{55 e^8}{12}
+\frac{329 e^{10}}{240} 
+
\mathcal{O}\left(e^{12}\right) 
\right] p^{-5} \log p
\\
& \hspace{3 ex}  +
\Bigg[
-\frac{1157}{15}-\frac{128 \gamma}{5}+\frac{677 \pi ^2}{512}-\frac{256}{5} \log 2 \\
& \hspace{8 ex} 
+
e^2  \left(-\frac{11141}{45}-\frac{592 \gamma}{15}+\frac{29665 \pi ^2}{3072}
+\frac{3248}{15} \log 2 -\frac{1458}{5} \log 3 \right) \\
& \hspace{8 ex} 
+
e^4  \left(\frac{247931}{360}+\frac{292 \gamma}{3}-\frac{89395 \pi ^2}{6144}
-\frac{64652}{15}\log 2+\frac{28431}{10}\log 3\right) \\
& \hspace{8 ex} 
+
e^6  \left(-\frac{52877}{180}-16 \gamma+\frac{3385 \pi ^2}{4096}+\frac{178288}{5} \log 2
-\frac{1994301}{160} \log 3 -\frac{1953125}{288} \log 5 \right) \\
& \hspace{8 ex} 
+
e^8  \left(-\frac{24619}{384}-\frac{55 \gamma}{6}+\frac{327115 \pi ^2}{196608}
-\frac{15967961}{90} \log 2 
+\frac{11332791}{1280} \log 3
+\frac{162109375}{2304} \log  5 \right) \\
& \hspace{8 ex} 
+
e^{10} \bigg(-\frac{1933}{3840}-\frac{329 \gamma}{120}+\frac{172697 \pi ^2}{393216}
+\frac{18046622551}{27000} \log 2+\frac{203860829079}{1024000} \log  3 \\
& \hspace{30 ex} 
-
\frac{74048828125}{221184} \log 5-\frac{678223072849}{9216000} \log 7\bigg)
+
\mathcal{O}\left(e^{12}\right) 
\Bigg]
p^{-5}
+
\mathcal{O}\left(p^{-6}\right) 
\end{split}	
\end{align}

\bibliography{EccentricPN}

\end{document}